\documentclass[prb, twocolumn,superscriptaddress,nofootinbib]{revtex4-2}
\makeatletter
\def\widetext@rule{}
\makeatother
\usepackage{epsfig}
\usepackage{amsmath}
\usepackage{graphicx}
\usepackage{color}
\usepackage[normalem]{ulem}
\usepackage[usenames,dvipsnames]{xcolor}
\usepackage[colorlinks=true,citecolor=Cerulean,linkcolor=RubineRed,urlcolor=Cerulean]{hyperref}
\usepackage{amsfonts,amssymb}
\usepackage[english]{babel}
\usepackage{hyperref}
\usepackage{lipsum}
\usepackage{graphics}
\usepackage{array}
\usepackage{tabu}
\usepackage{multirow}
\usepackage{bm}
\usepackage{float}
\usepackage{setspace}
\usepackage{IEEEtrantools} % Load the package

\newcommand{\be}{\begin{equation}}
\newcommand{\ee}{\end{equation}}
\newcommand{\ba}{\begin{eqnarray}}
\newcommand{\ea}{\end{eqnarray}}

\newcommand{\non}{\nonumber}
\newcommand{\bra}[1]{\langle #1|}
\newcommand{\ket}[1]{|#1\rangle}

\usepackage{lipsum}

\begin{document}

\title{Long-range interactions assisted shortcuts to adiabaticity and battery charging in open quantum  critical systems}
\author{Shishira Mahunta}
\affiliation{Department of Physical Sciences, IISER Berhampur, Berhampur 760003, India}
\author{Victor Mukherjee}
\affiliation{Department of Physical Sciences, IISER Berhampur, Berhampur 760003, India}

\begin{abstract} 

In this work we show that long-range interactions can be significantly beneficial for implementing shortcuts to adiabaticity (STA) in many-body open quantum critical systems driven out of equilibrium, as well as for charging quantum batteries in the presence of dissipation. In sharp contrast to short range interactions where passage through criticality may demand STA control with non-zero interactions between infinitely distant spins, using the example of a  Kitaev chain with long-range couplings, we find that the corresponding control may involve involve interaction strength with decays algebraically with distance. In case of non-unitary control, the advantage of long-range interactions manifest through reduction in the cost of STA. We further propose a modified STA technique aimed at charging a quantum battery in the presence of dissipation, in which case long-range interactions may enhance the resultant ergotropy. Our results establish long-range interactions as a valuable resource for quantum control, with direct implications for quantum technologies.
\end{abstract}
\maketitle
\section{Introduction}

Quantum systems driven out of equilibrium form a versatile area of research \cite{schaller14lecture}, owing to their fundamental nature, as well as their for potential applications in quantum technologies \cite{BhattacharjeeDutta20, campbell26roadmap} and condensed matter physics \cite{dziarmaga10dynamics}. Recent progress in experimental research has resulted in significant advancements in our ability to control the dynamics of quantum systems  for various tasks \cite{wurkner25identification}, including for preparation of desired quantum states  \cite{muller22one, ou25experimental}, for engineering novel phases of matter such as discrete \cite{sarkar26observation} and continuous \cite{phatthamon22observation} time crystals, and for designing quantum technologies \cite{koch23quantum}. However, quantum systems driven out of equilibrium are in general associated with non-adiabatic excitations \cite{dutta15quantum}. This can be detrimental, for example, for quantum annealing \cite{rajak22quantum}, or for modeling high-performing quantum technologies, including quantum computers \cite{hegade21shortcuts} and quantum engines \cite{hartmann19many}.  Consequently, studies on control in quantum systems, such as through shortcuts to adiabaticity (STA) \cite{Torrontegui13}, in order to reduce non-adiabatic excitations in driven quantum systems, are crucial for designing high-performing quantum technologies \cite{hartmann19many}. 

STA has been studied rigorously in closed quantum systems, where this technique may involve applying a control counterdiabatic (CD) Hamiltonian in order to ensure that the system remains in the instantaneous ground state of a time-dependent bare Hamiltonian at all times \cite{berry09transitionless}. On the other hand, studies on STA in open quantum systems have received far less attention. Yet, their importance cannot be overstated, owing to the ubiquitous presence of dissipation in nature, and also due to the relevance of open quantum systems driven out of equilibrium in quantum condensed matter physics \cite{schnell24dissipative} and quantum technologies \cite{cangemi24quantum}. STA in open quantum systems can entail designing a control protocol such that the system follows a particular target eigenstate of the Liouvillian \cite{vacanti14transitionless}. Alternatively, control can be desgined to ensure that the system remains in it's instantaneous thermal equilibrium corresponding to a particular temperature at all times, even in the presence of a time-dependent Hamiltonian \cite{alipour20shortcuts}. As recently reported in Ref. \cite{mahunta2025shortcuts}, in contrast to STA in unitary dynamics where this protocol may involve many-body control terms only close to criticality \cite{campo12assisted}, in the case of many-body open quantum systems  control of entropy may result in the STA protocol involving many-body interacting control terms even away from criticality. Therefore rigorous studies on practically implementable STA protocols in many-body open quantum systems are crucial for designing near-term quantum technologies. To this end, in this work we study STA in a long-range Kitaev (LRK) model driven through quantum critical points, in the presence of a Fermionic bath.
Our main motivation is to study whether the complexity of control  can be controlled by tuning the range of interaction present in the bare Hamilotnian. Notably, we show that in sharp contrast to STA in the presence of nearest neighbor interactions where the unitary control terms may involve infinite-range interactions at the critical point \cite{campo12assisted}, here long-range interactions (LRIs) result in a CD Hamiltonian  involving interaction strength which decays algebraically with distance. The same can contribute to reduction of the cost of STA in the case of non-unitary control.
We further use the protocol developed here to model a quantum battery in the presence of dissipation. Importantly, we  show that depending on the inherent critical features of the model, LRIs may assist in enhancing the ergotropy of the quantum battery.

Notably, LRIs have attracted significant attention in recent years \cite{defenu2019universal}, and have been used for studying STA in closed quantum systems \cite{campbell15shortcut}, for designing high-performing quantum engines \cite{solfanelli23quantum}, quantum batteries \cite{puri25floquet}, and have also found to be highly relevant for designing discrete \cite{russomanno17floquet} and continuous \cite{kozin19quantum} time crystals in the absence of disorder. 

The paper is organized as follows: in Sec. \ref{secModel} we present the model and dynamics; generic considerations are discussed in Sec. \ref{secGen}, followed by an analysis of the LRK model in Sec. \ref{secKitaev}. We study the unitary part of the STA protocol  in Sec. \ref{Unitary_CD_drive_QCP}, and the non-unitary control in Sec. \ref{STA_nonuni}. Analysis of the heat and power associated with STA are presented in Secs.  \ref{heat} and \ref{secpow}, respectively. We apply the protocol developed here to model a dissipation induced quantum battery with STA-enhanced ergotropy in Sec. \ref{secbat}. Finally we conclude in Sec. \ref{sec:conclusion}.

\section{Model and dynamics}
\label{secModel}
\subsection{STA in generic long-range interacting systems}
\label{secGen}

We now consider a quench across a quantum critical point in the presence of LRIs, with the interaction range controllable by a tuning parameter $\alpha$. In general, one can expect the corresponding energy gap $E(\alpha)$ to be a function of the parameter $\alpha$ as well, such that $E(\alpha) \to 0$ close to criticality \cite{sachdev99quantum}. This in turn leads to non-adiabatic excitations in quantum critical systems driven out of equilibrium \cite{dutta15quantum}, thus necessitating the implementation of STA control protocols \cite{campo12assisted}. In general, the STA control terms can be expected to be functions of $E(\alpha)$  (see below,  and also Refs. \cite{berry09transitionless} and \cite{alipour20shortcuts}), with the unitary control strength decreasing with increasing energy gaps \cite{takahashi13transitionless}. Consequently, we expect setups where one can tune $E(\alpha)$ by varying $\alpha$  can be ideal for implementation of practically realizable STA protocols, through appropriate choices of the parameter $\alpha$.

We exemplify the above using a detailed analysis of a LRK chain driven across quantum critical points. Importantly, a LRK chain is associated with two critical points with significantly different dispersion relations. In one case $E(\alpha)$ increases with increasing range of interactions (i.e., decreasing $\alpha$). As we discuss below, this allows us to implement STA with less demanding unitary control fields, through suitable choices of $\alpha$. On the other hand, the other critical point is associated with $E(\alpha)$ increasing with decreasing range of interactions (i.e., increasing $\alpha$), which can negate the advantages associated with LRIs.  
In addition,  we show that the above property of the LRK chain can also enable us to model a quantum battery with high ergotropy for longer ranges of interactions, through population inversion implemented by a modified STA protocol. 

\subsection{Long-range Kitaev chain}
\label{secKitaev}
We consider a LRK chain comprising spinless fermions on a one–dimensional lattice with long-range coupling. The Hamiltonian is given by \cite{dutta2017probing,defenu2019universal,vodola2014kitaev}
\begin{eqnarray}
H_0(t) &=& -\mathcal{J}\sum_{j}\left(c_j^{\dagger}c_{j+1}+\text{h.c.}\right)
-\mu(t)\sum_j\left(n_j-\frac{1}{2}\right) \nonumber \\
&& +\frac{1}{2}\sum_{i,j}\frac{1}{|i-j|^\alpha}
\left(c_i c_j+c_i^{\dagger}c_j^{\dagger}\right).
\label{bare_hamilotonian}
\end{eqnarray}
Here \(c_j^\dagger\) (\(c_j\)) are Fermionic creation (annihilation) operators at site \(j\), and \(n_j=c_j^\dagger c_j\) denotes the number operator. The parameter \(\mathcal{J}\) represents the nearest-neighbor hopping amplitude, while \(\mu(t)\) is the time-dependent chemical potential. We consider an interaction which decays algebraically with the distance between sites $i$ and $j$, with an exponent $\alpha > 0$; $\alpha \to \infty$ limit corresponds to nearest neighbor interactions in which case LRK chain reduces to the transverse field Ising chain \cite{vodola2014kitaev, vodola2016long}, while smaller values of $\alpha$ corresponding to enhancement in the range of interactions. 
Introducing the momentum-space Fermionic operators 
\[
c_j=\frac{1}{\sqrt{L}}\sum_{k} e^{-ikj} c_k 
\]
with antiperiodic boundary conditions \(c_{j+L}=-c_j\), the Hamiltonian decouples into independent momentum sectors. In the Nambu basis:  \(\Psi_k^\dagger=(c_k^\dagger,c_{-k})\), the Hamiltonian assumes the Bogoliubov–de Gennes form, 
\begin{eqnarray}
H_0(t) &=& \sum_k \Psi_k^\dagger  H_{0k} \Psi_k,
\label{eq:free_fermionic}
\end{eqnarray}
with $H_{0k}(t)  = \left[(\mu+2\mathcal{J}\cos k)\sigma^z + i\,\, f_\alpha(k)\,\sigma^x \right]$, and  \(f_\alpha(k)\) denotes the Fourier transform of the long-range pairing function:
\begin{equation}
f_{\alpha}(k)
= \frac{1}{2i}\sum_{l=1}^{\infty}\frac{e^{ilk}-e^{-ilk}}{l^\alpha}
= \sum_{l=1}^{\infty}\frac{\sin(lk)}{l^\alpha}.
\label{eq:f_alpha}
\end{equation}
In the occupation-number basis 
\(\{ \ket{0_k,0_{-k}},\ket{0_k,1_{-k}},\ket{1_k,0_{-k}},\ket{1_k,1_{-k}}\}\) one can write 
\begin{align}
%H_0(t) &= \sum_k H_{0k}(t), \nonumber \\
 H_{0k}(t) &= \frac{1}{2}
\begin{bmatrix}
(2\mathcal{J}\cos k+\mu) & 0 & 0 & i f_\alpha(k) \\
0 & 0 & 0 & 0 \\
0 & 0 & 0 & 0 \\
-i f_\alpha(k) & 0 & 0 & -(2\mathcal{J}\cos k+\mu)
\end{bmatrix}.
\label{kthHamiltonian}
\end{align}
We note that while the unitary dynamics couples only the even-parity subspace, the inclusion of dissipation necessitates working in the full four-dimensional basis \cite{keck17dissipation, bandyopadhyay18exploring}.
Diagonalizing Eq.~(\ref{kthHamiltonian}) via a Bogoliubov transformation yields the quasiparticle spectrum
\begin{equation}
\tilde{E}_k^{(1,4)}=
\pm\sqrt{(2\mathcal{J}\cos k+\mu)^2+(f_\alpha(k))^2} = \pm E_k(\alpha),
\label{eq:spectrum}
\end{equation}
while the remaining two eigenvalues satisfy \(\tilde E_k^{(2)}=\tilde E_k^{(3)}=0\). 
In order to have non diverging energy spectrum in k-space, we only consider $\alpha > 1$ \cite{dutta2017probing, defenu2019universal}.
As seen from Eq. \eqref{eq:spectrum}, the energy gap vanishes for the modes $k = \pi$ and $k = 0$ at the critical points  $ \mu = 2\mathcal{J}$ and $ \mu = -2\mathcal{J}$, respectively \cite{kitaev2001unpaired}. 
Throughout this work we set \(\mathcal{J}=1/2\), the Planck constant $\hbar$ and the Boltzmann constant $k_B$ are set to unity, and the chemical potential $\mu(t)=\frac{t}{\tau}$ is varied linearly in time 
with a quench rate $\tau$.

We first focus on the long-wavelength behavior of the quasiparticle dispersion $E_k(\alpha)$ near the critical points $\mu=\pm1$. Close to $\mu=-1$, the Taylor expansion of $E_k(\alpha)$ around $k \to 0$ leads to, 
\begin{equation}
E_k(\alpha, \mu\to-1)
\approx
\begin{cases}
\sqrt{\epsilon_-^2+A_\alpha^2k^{2(\alpha-1)}}, & 1<\alpha<2, \\[0.2cm]
\sqrt{\epsilon_-^2+B_\alpha^2k^2}, & \alpha>2,
\end{cases}
\label{eq:Ek_minus_one}
\end{equation}
where $\epsilon_{-}=\mu+1$ denotes the distance from the criticality, 
$A_\alpha=2\left|\cos\left(\frac{\pi\alpha}{2}\right)\Gamma(1-\alpha)\right|$ and 
$B_\alpha=\zeta(\alpha-1)$. Here, $\Gamma(x)$ and $\zeta(x)$ denote the Gamma and Riemann zeta functions, respectively.
Owing to the change in the form of  $E_k(\alpha)$ at $\alpha = 2$ (see Eq. \eqref{eq:Ek_minus_one}), and following Refs. \cite{dutta2017probing,defenu2019universal},  below we refer to $1<\alpha<2$ as the long-range regime, while $\alpha > 2$ constitute the short-range regime.   $\alpha = 2$ denotes the boundary between short and long-range regimes, and is associated with a logarithmic correction  (see Appendix~\ref{App:A} for details).

Close to $\mu=+1$, we get
\begin{equation}
E_k(\alpha, \mu\to+1)
\approx
\sqrt{\epsilon_+^2+\tilde B_\alpha^2q^2},
\label{eq:Ek_plus_one}
\end{equation}
 for both long-range and short-range regimes. Here $\epsilon_+=\mu-1$,  $q=\pi-k$ and $\tilde B_\alpha=\eta(\alpha-1)$, with $\eta(s)$ denoting the Dirichlet eta function. Further, in sharp contrast to the case of $\mu \to -1$ where $E_k(\alpha, \mu \to -1)$ increases with decreasing $\alpha$, here $\tilde B_\alpha$, and hence $E_k(\alpha, \mu\to+1)$,  monotonically increases  with increasing $\alpha$ for $\alpha\ge 1 $ and approaches $1$ as $\alpha\to \infty$. This suggests qualitative differences between the dynamics of system close to the two critical points, as we verify below through rigorous analysis. 

We assume that the setup is coupled to a Fermionic bath \cite{keck17dissipation}, such that an adiabatic dynamics with $\tau \to \infty$ results in each momentum mode remaining in its instantaneous thermal state
\ba
 \rho_k^{\mathrm{th}}(t) &=& \sum_n \lambda_n^k(t) \ket{n^k_t}\bra{n^k_t} \non\\
&=&\begin{bmatrix}
\frac{e^{-\beta E_k(\alpha)}}{\mathcal{Z}_k} & 0 & 0 & 0 \\
0 & \frac{1}{\mathcal{Z}_k} & 0 & 0 \\
0 & 0 & \frac{1}{\mathcal{Z}_k} & 0 \\
0 & 0 & 0 & \frac{e^{\beta E_k(\alpha)}}{\mathcal{Z}_k}
\end{bmatrix}
\label{trajectory}
\ea
at any time $t$. 
Here \(\ket{n^k_t}\) are the instantaneous eigenstates of \(H_{0k}(t)\), 
 \(\lambda_n^k = e^{-\beta \tilde{E}_n(k)}/\mathcal{Z}_k\), and $\mathcal{Z}_k = 2 + e^{-\beta E_k} + e^{\beta E_k}$ denotes the partition function (see Eqs. \eqref{kthHamiltonian} and \eqref{eq:spectrum}).
 
In general a quench with a finite $\tau$ would result in the setup going out of thermal equilibrium with the Fermionic bath. However, here we consider control in the form of STA in order 
to enforce the trajectory (\ref{trajectory}), even in the presence of finite $\tau$. To this end, following Refs. \cite{alipour20shortcuts} and \cite{mahunta2025shortcuts}, we assume each mode of the system evolves according to a  Lindblad type master equation
\ba \label{eq:master}
\dot{\rho}_k(t) &=& \mathcal{L}^{k}[\rho(t)] = -i\left[H^k_{\rm STA}, \rho_k(t)\right] \non\\
&+& \sum_{m,n}\gamma^{(k)}_{mn} \left(A^{(k)}_{mn}\rho_k A_{mn}^{(k)\dagger} -\frac{1}{2}\{A_{mn}^{(k)\dagger}A^{(k)}_{mn},\rho_k \}\right)\non\\
H^k_{\rm STA} &=&  H_{0k} + H^k_{CD}.
\ea
Here
\ba
%H^k_{\rm STA} &=&  H_{0k} + H^k_{CD} \non\\
H_{CD}^k(t) = \mathrm{i} \hbar \sum_n\left(\left|\partial_t n^k_t\right\rangle\left\langle n^k_t\left|-\left\langle n^k_t \mid \partial_t n^k_t \right\rangle\right| n^k_t\right\rangle\langle n^k_t|\right) \non\\
\label{eq:Hcd}
\ea
is the counterdiabatic Hamiltonian with respect to $H_{0k}(t)$  \cite{berry09transitionless, campo12assisted}. The control Lindblad operators in the instantaneous eigen
basis of $H_{0k}(t)$  are given by 
\ba \label{eq:gamma}
A^k_{mn}(t) = \ket{m^k_t}\bra{n^k_t},
\ea
 and 
\ba
\gamma_{mn}^k(t) = \frac{\dot{\lambda}^k_{m}(t)}{r \lambda^k_n(t)}
\label{eq:gamma2}
\ea
denotes the rate of transition from the $\ket{n_t^k}\bra{n_t^k}$ state to the state $\ket{m_t^k}\bra{m_t^k}$ for the mode $k$, with
$r$ being the rank of $\rho(t)$.
 %%%%%%%%%%%%%%%%%%%%%%%%%%%%%%%%%%%%
$\ket{m^k_t},~\ket{n^k_t}$ denote any of the four instantaneous eigenstates for the $k$-th mode, given by 
\ba 
\ket{\xi^k_1(t)} &=& \frac{\phi_t \ket{1_k, 1_{-k}} + \ket{0_k, 0_{-k}}}{\sqrt{|\phi_t|^2 + 1}}, \non\\
\ket{\xi^k_2(t)} &=& \ket{1_k, 0_{-k}},\non\\
\ket{\xi^k_3(t)} &=&  \ket{0_k, 1_{-k}},\non\\
\ket{\xi^k_4(t)} &=& \frac{\theta_t \ket{1_k, 1_{-k}} + \ket{0_k, 0_{-k}}}{\sqrt{|\theta_t|^2 + 1}}, 
\label{levelsk}
\ea 
where 
\ba \label{Eq:population}
\phi^{(k)}_t &=& \frac{\left(\mu + \cos k - E_k\right)}{f_{\alpha}(k)},  \non\\
\theta^{(k)}_t &=& \frac{\left(\mu + \cos k + E_k\right)}{f_{\alpha}(k)}. 
\ea
Here, $\ket{\xi^k_1(t)}$ is the ground state, $\ket{\xi^k_2(t)}$ and $\ket{\xi^k_3(t)}$ are degenerate second excited states, and $\ket{\xi^k_4(t)}$ is the highest excited state. 
%We note that,  $A^k_{mn}(t) = \ket{m^k_t}\bra{n^k_t}$ are the controlled Lindblad operators in instantaneous eigen basis of $H_{0k}(t)$, and corresponding coupling rates,  $\gamma_{mn}^{(k)}$ denotes the rate of transition from the $\ket{n^k}\bra{n^k}$ state to the state $\ket{m^k}\bra{m^k}$, for the mode $k$.
Further, the free Fermionic nature of the system allow us to write
\ba
\rho(t) = \bigotimes_k \rho_k(t),
\ea
such that
\ba
\dot{\rho}(t) &=& \dot{\rho}_{k1}\bigotimes_{k \neq k_1} \rho_k + \rho_{k1} \otimes \dot{\rho}_{k2} \bigotimes_{k \neq k_1, k_2}\rho_k + \hdots \non\\
&=& \sum_k \mathcal{L}_k \left[\rho\right] = \mathcal{L}\left[\rho \right].
\ea
Below we examine separately the role of LRIs in determining the unitary CD Hamiltonian \(H_{CD}^k\), which ensures adiabatic evolution across the critical point in the closed system, and the dissipative control terms that, together with \(H_{CD}^k\),  constrains the dynamics to follow the target trajectory Eq.~(\ref{trajectory}).

\section{CD drive across a QCP with LRIs}
\label{Unitary_CD_drive_QCP}
Now we investigate the effect of LRIs on   \(H_{CD}^k\) (Cf. Eq. \eqref{eq:Hcd}). 
Following the standard construction
of transitionless driving for free-fermionic systems exhibiting quantum
phase transitions \cite{berry09transitionless, campo12assisted}, the corresponding
$H_{CD}^k$ in the momentum space can be written as (see Appendix \ref{App:CD_LRK})
\begin{equation}
H_{\rm CD}^k(t)
=
-\dot{\mu}(t)\sum_k \frac{1}{2}
\frac{f_\alpha(k)}
{(\mu+\cos k)^2+f_\alpha^2(k)}
\,\Psi_k^\dagger \sigma^y \Psi_k .
\label{eq:HCD_momentum}
\end{equation}
Fourier transforming Eq.~\eqref{eq:HCD_momentum} to real space yields
\begin{eqnarray}
    H_{\rm CD}(t) &=& \dot{\mu}(t)\sum_{m\ge 1} h_m(\mu,\alpha)\,\hat O_m,~~~\text{where}
 \non \\
\quad \hat O_m &=& \sum_j i\left(c_j c_{j+m}-c_{j+m}^\dagger c_j^\dagger\right)\non\\
h_m(\mu,\alpha) &=& \frac{1}{4\pi}\int_0^\pi dk\;\frac{f_\alpha(k)\sin(mk)}{(\mu+\cos k)^2+f_\alpha^2(k)}.
\label{eq:Cd_qcp}
\end{eqnarray}
Here $\hat{O}_m$ denotes the operators acting on sites separated by a distance $m$, while $h_m(\mu,\alpha)$ denotes the strength of the associated CD field. 
%%%%%%%%%%%%%%%%%%%%%%%
In the short-range limit \(\alpha\to\infty\), the pairing function reduces to
$f_\alpha(k)\to \sin k,$ and Eq.~\eqref{eq:HCD_momentum} reproduces the nearest-neighbor transverse-field Ising model results derived in Ref.~\cite{campo12assisted}. In that limit, the CD coefficients decay exponentially with distance,
$h_m(\mu)\sim e^{-m/\xi(\mu)}$
with \(\xi(\mu)\) being the  correlation length of transverse-field Ising model. However, as we show below, the long-range structure of \(f_\alpha(k)\) may  lead to a qualitatively different spatial structure of the CD field, especially near criticality.

% %%%%%%%%%%%%%%%
\subsection{CD drive across the critical point \texorpdfstring{$\mu = -1$}{mu\_c = -1}}
\label{subsec:SR_offcritical}

\subsubsection{Long-range regime \texorpdfstring{($1 < \alpha < 2$)}{(1 < alpha < 2)}}
We first analyze the behavior of the CD coupling coefficient $h_m(\epsilon_-, \alpha)$ near the critical point  $\mu=-1$ in the long range regime ($1< \alpha<2$). 
Using Eqs. (\ref{eq:Ek_minus_one}) and (\ref{eq:Cd_qcp}), and the Taylor series expansion of the of the  function $f_\alpha(k)$ (see Appendix \ref{App:A}), one gets 
\begin{equation}\label{eq:h_r_LR}
    h_m(\epsilon_-, \alpha) \approx \frac{1}{4\pi} \int_0^\pi \frac{A_\alpha k^{\alpha-1}}{\epsilon_-^2 + A_\alpha^2 k^{2(\alpha-1)}}\, \sin(mk)\,  dk.
\end{equation}

%%%%%%%%%%%%%%%%%%%%%%%%%%%%%%%%%%%%%%%%%%%%%%%%%%%%%%%%%%%%%%%%%%%%%%%

{\textbf{(i) Away from criticality $(\epsilon_- \neq 0)$:} Introducing a scaling variable $y=m/m_\epsilon,~m_{\epsilon} = \left(A_\alpha / |\epsilon_-|\right)^{\!1/(\alpha-1)}$ and performing the 
change of variable $k = x/m_{\epsilon}$, Eq.~\eqref{eq:h_r_LR} can be written 
in the form
\begin{align}
h_m(\epsilon_-,\alpha)
&\approx
\frac{1}{4\pi}\,
A_\alpha^{-1/(\alpha-1)}
|\epsilon_-|^{(2-\alpha)/(\alpha-1)}
\,
\Phi_\alpha\!\left(\frac{m}{m_\epsilon}\right),
\non \\ \text{where} \quad 
\Phi_\alpha (y) &= \int_0^\infty dx\;
\frac{x^{\alpha-1}}{1+x^{2(\alpha-1)}} \sin(yx).  
\label{eq:hm_scaling_LR}
\end{align}
The asymptotic behavior of $h_m$ follows from the limits of the scaling function $\Phi_\alpha(y) $. 
For distances much longer than the crossover scale ($y \gg 1$ or $m \gg m_\epsilon $), the factor $\sin(yx)$ becomes rapidly
oscillatory. As a result, the contributions from large $x$ cancel out,
and the integral is dominated by the small-$x$ region, resulting in $\Phi_\alpha(y) \approx \int_0^\infty dx \,\, \, x^{\alpha-1} \sin(yx)$.
Therefore we finally get 
\begin{equation}
\lim_{m \gg m_{\epsilon}} h_m(\epsilon_{-},\alpha)
\approx
\frac{A_\alpha\,\Gamma(\alpha)\sin(\pi\alpha/2)}{4\pi}
\;
\frac{1}{\epsilon_{-}^2}
\
\frac{1}{m^{\alpha}} .
\label{eq:hm_LR_offcritical}
\end{equation}
On the other hand, for the distance much smaller than the crossover scale ($y\ll 1$ or $m \ll m_\epsilon$), the dominant contribution of the scaling function, $\Phi_\alpha(y)$ comes from the large $x$ region. Therefore the integral in Eq. \eqref{eq:hm_scaling_LR} can be approximated as $ \sim \int_0 ^ \infty dx \, x^{1-\alpha} \sin(yx) $ which reduces to, 
\begin{eqnarray}
\lim_{m \ll m_{\epsilon}}  h_m(\epsilon_{-},\alpha)&\approx& \frac{C(\alpha)}{m^{2-\alpha}},
\label{msmall}
\end{eqnarray}
Where $C(\alpha) = \frac{1}{4\pi}(1-\alpha)\tan\!\left(\frac{\pi\alpha}{2}\right)$ (see also Appendix \ref{App:CD_LRK}). %is given in Eq.\eqref{eq:hm_LR_critical}. 
We verify the above analytical forms (Eqs. \eqref{eq:hm_LR_offcritical} and \eqref{msmall}) with numerically obtained values of $h_m$ in Fig. \ref{fig:hm_vs_m_mu_minus_one_LR}; as seen in the main panel of   Figs. \ref{fig:hm_vs_m_mu_minus_one_LR}(a), the crossover between the two scaling forms $h_m \sim m^{-(2 - \alpha)}$ for $m \ll m_{\epsilon}$ and $h \sim m^{-\alpha}$ for $m \gg m_{\epsilon}$ occurs at $m \sim m_{\epsilon}$.

%%%%%%%%%%%%%%%%%%%%%%%%%%%%%%%%%%%%%%%%%%%%%%%%%%%%%%%%%%%%%%%%%%%%%%%%%%%%%%%%%%%
{\textbf{(ii) At the  criticality $(\epsilon_- =  0)$:} At the critical point  $\mu =-1$ we have (see Eq. (\ref{eq:h_r_LR}))  
\begin{equation*}
   h_m(\epsilon_- = 0, \alpha) \approx \frac{1}{4\pi A_\alpha}\int_0^\pi  k^{(1- \alpha)}\sin(mk) \, dk. 
\end{equation*}
In the limit of $m \gg 1$, we get (see Appendix \ref{App:CD_LRK} for a detailed derivation) back the result Eq. \eqref{msmall}:
\begin{equation}
 \lim_{m \gg 1} h_m(\epsilon_- = 0, \alpha)
\approx
\frac{C(\alpha)}{m^{2-\alpha}}.
\label{eq:hm_LR_critical}
\end{equation}
In sharp contrast to the case of nearest neighbor interactions  where introduction of STA demands infinite-range control terms at the critical point (see Ref. \cite{campo12assisted}), here the CD couplings $h_m$  decay algebraically with an exponent
continuously tunable by $\alpha$, thus emphasizing the significant advantage offered by LRIs, as also verified numerically in the inset of the Fig. \ref{fig:hm_vs_m_mu_minus_one_LR}(a). This is a direct consequence of the energy gap closing sublinearly in the long-range regime for $\mu \to -1$.
%Notably, while infinite-range interactions may be impossible to implement in practical scenarios, here LRIs may allow us to implement exact STA protocols, even for passages through criticality.

\subsubsection{Short-range regime \texorpdfstring{($\alpha > 2$)}{(alpha > 2)}}
\label{subsec:SR_regime}

We now analyze the behavior of the CD coupling $h_m(\epsilon_-,\alpha)$ in the
short-range regime $\alpha>2$. 
Following Eq.~\eqref{eq:Cd_qcp}, in this regime  one gets
\begin{equation}
h_m(\epsilon_-,\alpha>2)
\approx
\frac{1}{4\pi}
\int_0^\pi
\frac{B_\alpha k\,\sin(mk)}
{\epsilon_-^2+B_\alpha^2 k^2}\, dk .
\label{hmalgtr2}
\end{equation}
As shown in Appendix \ref{App:CD_LRK}, in the limit of $m>>1$ Eq. \eqref{hmalgtr2} finally leads us to
\begin{equation}
h_m(\epsilon_-,\alpha>2)
\approx
\frac{1}{8\,\zeta(\alpha-1)}
\exp\!\left[
-\frac{|\epsilon_{-}|}{\zeta(\alpha-1)}\, m
\right].
\label{eq:hr_SR_minus}
\end{equation}
Thus, away from the critical point the CD couplings are exponentially
localized in real space, with a decay length 
$
\xi_{\mathrm{CD}}^{(-)}
=
\frac{\zeta(\alpha-1)}{|\epsilon_{-}|}.$ At the critical point ($\epsilon_{-}=0$), the exponential factor
in Eq.~\eqref{eq:hr_SR_minus} approaches unity and the CD couplings
saturate to the constant value
\begin{equation}
h_m(\epsilon_-=0,\alpha>2)
\approx
\frac{1}{8\,\zeta(\alpha-1)},
\label{eq:h_m_SR_critical}
\end{equation}
analgous to that seen earlier for transverse Ising model with nearest neighbor interactions in the limit of $\alpha \to \infty$ \cite{campo12assisted}.
The above analytical results are verified numerically in Fig.~\ref{fig:hm_vs_m_mu_minus_one_LR}(b).

\begin{figure*}
    \centering
    \includegraphics[width=1\linewidth]{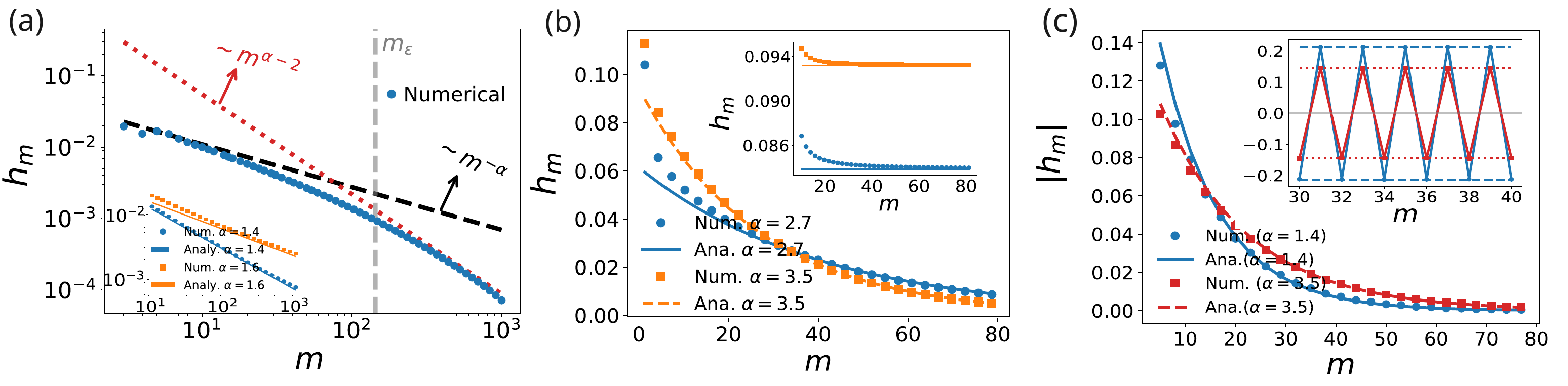}
    \caption{(a) Comparison between the numerically evaluated CD coupling $h_m(\mu,\alpha)$ and the analytical predictions in the long-range regime. The main panel shows the behavior close to criticality $\mu = -1$ ($\epsilon_-=0.3$, $\alpha =1.4$); the CD coupling exhibits a crossover from  a detuned asymptotic decay described by Eq.~\eqref{eq:hm_LR_offcritical} to a  power-law behavior Eq.~\ref{msmall}, controlled by the characteristic length scale $m_\epsilon=(A_\alpha/|\epsilon|)^{1/(\alpha-1)}$. The inset shows the critical scaling at $\epsilon_-=0$ for $\alpha=1.4$ and $\alpha=1.6$ (see Eq.~(\ref{eq:hm_LR_critical})). (b) Comparison between numerical and analytical results for the CD coupling in the short-range regime, showing exponentially localized behavior close criticality $\mu = -1$ (see Eq.~\ref{hmalgtr2}) (main panel) and the corresponding critical behavior at $\\epsilon_0 = 0$ (see Eq.~\ref{eq:h_m_SR_critical}) (inset). 
    (c) Comparison between numerical and analytical results for the CD coupling near $\mu=+1$, showing the behavior away from criticality (see Eq.~\ref{eq:hr_SR_plus}) (main panel) and at criticality $\epsilon_+ = 0$ (see Eq.~\ref{eq:h_m_critical_mu_one}) (inset). Numerical results are obtained from direct integration of Eq.~\eqref{eq:Cd_qcp}.} 
    \label{fig:hm_vs_m_mu_minus_one_LR}
\end{figure*}

\subsection{CD drive across the critical point \texorpdfstring{$\mu=+1$}{mu = +1}}

We now analyze the behavior of the CD coupling near the second critical
point $\mu=+1$. 
Repeating
the same steps as in  Section \ref{subsec:SR_regime} yields 
a staggered modulation of the CD coupling:
\begin{equation}
h_m(\epsilon_+,\alpha)
\approx
(-1)^{m+1}
\frac{1}{8\,\eta(\alpha-1)}
\exp\!\left[
-\frac{|\epsilon_+|}{\eta(\alpha-1)}\,m
\right],
\label{eq:hr_SR_plus}
\end{equation}
where we have used $\sin[m(\pi-q)] = (-1)^{m+1}\sin(mq)$ and $\epsilon_+ = \mu -1$ (see Appendix \ref{App:CD_LRK} for more details).
Analogous to that seen close to the $\mu = -1$ criticality, the magnitude of the CD couplings are exponentially localized in real space in this case as well,
with a $m~$ dependent  sign.
The corresponding decay length is $
\xi_{\mathrm{CD}}^{(+)}
=
\frac{\eta(\alpha-1)}{|\mu-1|}. $
\begin{table*}[t]
\centering
\caption{Behavior of the CD coupling strength $h_m(\mu,\alpha)$ in the long-range (LR) ($1<\alpha<2$) and short-range (SR) ($\alpha>2$) regimes.}
\renewcommand{\arraystretch}{1.4}
\begin{tabular}{|c|c|c|c|}
\hline
\textbf{Critical point} & \textbf{Regime} & \textbf{At criticality ($\epsilon_{\pm}=0$)} & \textbf{Away from criticality ($\epsilon_{\pm}\neq0$)} \\
\hline

\multirow{2}{*}{$\mu=-1$}
& LR ($1<\alpha<2$)
& $h_m \sim m^{-(2-\alpha)}$
&
\begin{tabular}{c}
$m\ll m_\epsilon:\quad h_m \sim m^{-(2-\alpha)}$ \\
$m\gg m_\epsilon:\quad h_m \sim m^{-\alpha}$
\end{tabular}
\\
\cline{2-4}

& SR ($\alpha>2$)
& $h_m \rightarrow \text{constant}$
& $h_m \sim e^{-m/\xi_{\rm CD}^{(-)}}$
\\

\hline

$\mu=+1$
& LR  and SR
& $h_m \sim (-1)^{m+1}$
& $h_m \sim (-1)^{m+1}e^{-m/\xi_{\rm CD}^{(+)}}$
\\

\hline
\end{tabular}
\label{tab:CD_summary}
\end{table*}
At the critical point ($\epsilon_+  = 0$), the exponential factor in
Eq.~\eqref{eq:hr_SR_plus} becomes unity and the CD couplings
saturate to the constant magnitude
\begin{equation}
h_m(\mu=+1)
\approx
(-1)^{m+1}\,
\frac{1}{8\,\eta(\alpha-1)} .
\label{eq:h_m_critical_mu_one}
\end{equation}
The above  analytical results Eqs. \eqref{eq:hr_SR_plus} and \eqref{eq:h_m_critical_mu_one} are verified numerically in Fig. \ref{fig:hm_vs_m_mu_minus_one_LR}(c).
\begin{figure}
    \centering
    \includegraphics[width=1\linewidth]{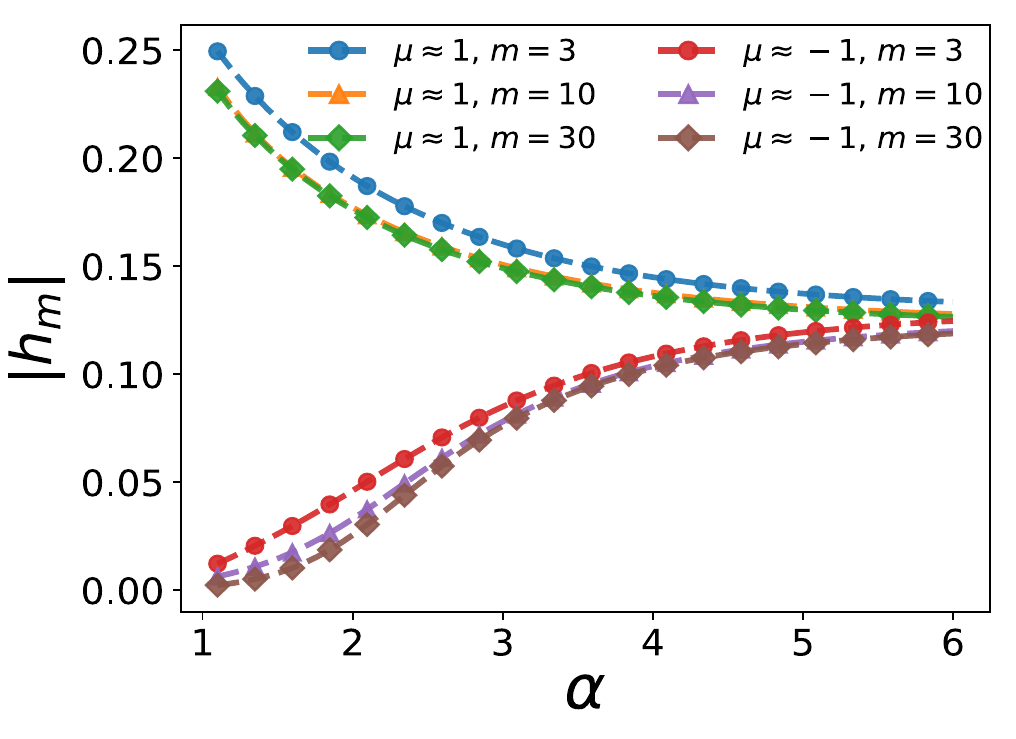}
    \caption{ Dependence of $h_m(\mu,\alpha)$, obtained through numerical integration of Eq. \eqref{eq:Cd_qcp}, on the interaction exponent $\alpha$ near $\mu=\pm1$. Here $\epsilon_{\pm}  = 0.1$.}
    \label{fig:hm_critical_vs_alpha}
\end{figure}

\subsection{Variation of \texorpdfstring{$h_m$}{h\_m} with range of interaction \texorpdfstring{$\alpha$}{alpha}}

We plot  the magnitude of the CD field $h_m$ (see Eq.~\eqref{eq:Cd_qcp}) close to the critical points $\mu = \pm 1$, as a function of the interaction exponent $\alpha$, in Fig. ~\ref{fig:hm_critical_vs_alpha}.
Notably,  $|h_m|$ decreases as the system moves from the
short-range regime ($\alpha>2$) toward the long-range regime
($1<\alpha<2$) for $\mu = -1$. This stems from the enhancement of $E_k(\alpha)$ with decrease in $\alpha$ for $k \to 0$, thus showing the advantage of LRIs for suppressing non-adiabatic excitations in the this regime. 
On the other hand, the enhancement of $E_k(\alpha)$ with increasing $\alpha$ for $k \to \pi$ results in the opposite trend being observed near the second critical
point $\mu=+1$, where short-range interactions require stronger control fields.

We summarize the behavior of the counterdiabtic coupling across the critical points, $\mu = \pm 1$ in long and short regimes in Table-\ref{tab:CD_summary}.

}

\section{STA in the presence of dissipation}
\label{STA_nonuni}

In this section we investigate how LRIs modify the
CD control associated with the dissipative dynamics.
The system evolves according to the  Lindblad master equation
[Eq.~\eqref{eq:master}], where the dissipative STA protocol is implemented
through time-dependent Lindblad operators $A_{mn}^{(k)}(t)$ and the
corresponding transition rates $\gamma_{mn}^{(k)}(t)$.
The explicit expressions for $A_{mn}^{(k)}$ and $\gamma_{mn}^{(k)}$
are summarized in Appendix~\ref{app:Lindblad_operators_and_rates}. The forms  of $\gamma_{mn}^{(k)}$'s remain similar to that evaluated for the nearest neighbor interaction case in Ref. \cite{mahunta2025shortcuts},  albeit with the introduction of the parameter $f_{\alpha}$. As we discuss below,  this in turn results in non-trivial dependence of the control terms on the range of interactions.

\subsection{Impact of LRIs on the dissipative control strength}

We analyze the effect of the long-range exponent $\alpha$ on transition rates $\gamma_{mn}^{(k)}$ across the critical points $\mu=\pm1$, focusing on different temperature regimes.  
We first consider $\mu=-1$, where the gap closes at $k=0$ and the spectrum is non-linear. 

\textbf{\textit{(i) Low-temperature regime:}}
The variation of $\gamma_{mn}^{(k)}$'s in the low temperature regime are shown in Figs. \ref{fig:gamma_low_highTemp}a - \ref{fig:gamma_low_highTemp}d. In the limit of $\beta E_k \gg 1$, the system remains close to its instantaneous ground state. Consequently, excitation rates from the ground state to higher excited states (e.g., $\gamma_{21}, \gamma_{41},$) are exponentially suppressed for large $\beta$, except in the vicinity of the gap-closing point ($E_k \to 0$) (see Figs.~\ref{fig:gamma_low_highTemp}(a), \ref{fig:gamma_low_highTemp}(b) and Eq. \eqref{eq:gamk}).  This suppression is more pronounced for small $\alpha$, owing to larger values of $E_{k}(1 < \alpha < 2, \mu = -1) \propto k^{\left(\alpha - 1\right)}$ in this regime (see also Appendix \ref{App:A}).
In contrast, de-excitation rates from higher energy levels (e.g., $\gamma_{12}, \gamma_{14} $) scale as
$\gamma_{mn}^{(k)} \propto \frac{e^{\beta E_k}}{E_k},$ 
(see Eq.~(\ref{eq:gamk_low_temp})), and thus become strongly enhanced as the system relaxes toward the ground state.

\textbf{\textit{(ii) High-temperature regime:}} 

In the high-temperature limit of $\beta E_k \to \infty$, the system approaches a maximally mixed state with nearly equal population of all energy levels, resulting in small values of $\gamma_{mn}$ 
as seen in Figs.~\ref{fig:gamma_low_highTemp}(e) and \ref{fig:gamma_low_highTemp}(f).  Several dissipative control terms  $\{viz.~\gamma^{(k)}_{12}, \gamma^{(k)}_{23}, \gamma^{(k)}_{21}, \gamma^{(k)}_{41}, \gamma^{(k)}_{42}\} \sim 1/E_k(\alpha)$, thereby decreasing with increasing range of interactions (i.e., decreasing $\alpha$). On the other hand, the remaining rates become largely insensitive to $\alpha$ in this limit  (see Appendix~\ref{app:Lindblad_operators_and_rates}).

We now turn to the critical point $\mu=+1$, where the gap closes linearly for  $k \to \pi$, $\forall~\alpha$, both in high as well as low temperature regimes. Unlike the $\mu=-1$ case, in this case the low-energy spectrum increases as the system crosses over from long-range to short-range interactions (i.e., with increasing $\alpha$).  
As a consequence, transition rates that scale as $\gamma^{(k)}_{mn} \sim 1/E_k(\alpha)$ (eg. $\gamma^{(k)}_{41}, \gamma^{(k)}_{31} = \gamma^{(k)}_{21}, \gamma^{(k)}_{23} = \gamma^{(k)}_{32}$) decrease with increasing $\alpha$, thereby indicating the requirement of stronger rates of dissipative control for longer interaction ranges (see  Fig.(~\ref{fig:gamma_low_highTemp})); the remaining $\gamma$'s become insensitive to $\alpha$, as also seen for $\mu \to -1$. Furthermore, analogous to the $\mu = -1$ case, high temperatures correspond to small values of $\gamma^{(k)}_{mn}$'s, owing to thermal mixing.

\begin{figure*}
    \centering
    % \vspace{-1em} % reduces top white space
    \includegraphics[width=1\linewidth]{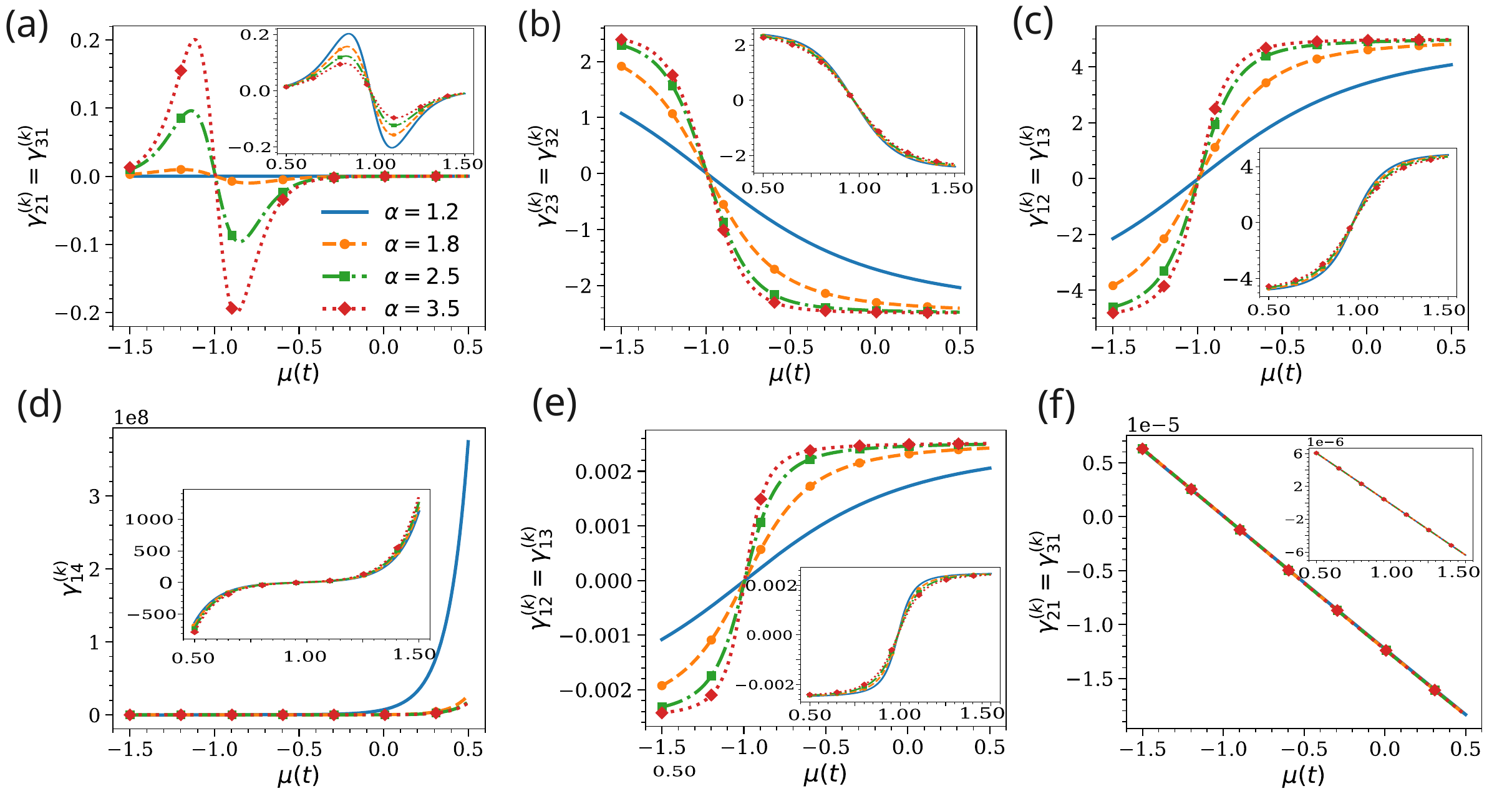}
    \vspace{-1em} % reduces bottom white space
    \caption{(a)–(d) Main panels show the transition rates $\gamma^{(k)}_{mn}(t) $ as a function of the time-dependent chemical potential $\mu(t)$ driven across the critical point $\mu =-1$, for $k = 0.1) $. The insets display  $\gamma^{(k)}_{mn}(t)$ with $\mu(t)$ driven across the critical point $\mu =+1$, for  $k = 3.0$. 
    Results are presented for different values of the LRI exponent $\alpha$ in low temperature regime $\beta=10$. (e),(f) Same as above, but in the high-temperature regime $\beta = 0.01$.}
    \label{fig:gamma_low_highTemp}
\end{figure*}

\section{Power and heat current under STA protocol}
\label{secPower_Heat}

In this section we investigate the effect  of the LRIs on the  energetic  costs associated with the STA protocol, quantified  by the heat and  power generated in this process. 

% as determined by the corresponding heat current $\mathcal{I}_{STA}$ and power dissipated  $\mathcal{P}_{STA}$.  
\subsection{Heat Current}
\label{heat}
Following Refs.~\cite{spohn78entropy, mahunta2025shortcuts, alipour20shortcuts},
the heat current under the STA protocol is defined as
$\mathcal{I}_{\mathrm{STA}} = \mathrm{Tr}(\dot{\rho} H_{\mathrm{STA}})$,
where $H_{\mathrm{STA}}=H_0+H_{\mathrm{CD}}$ is the total Hamiltonian.
Using the  master equation \eqref{eq:master}
\cite{alicki79The}, the heat current becomes
\ba
\mathcal{I}_{\mathrm{STA}} &=& \mathrm{Tr}\!\left(-i[\rho,H_{\mathrm{STA}}]H_{\mathrm{STA}}+\mathcal{L}(\rho)H_{\mathrm{STA}}\right)\non\\
&=& \mathcal{I}_0+\mathcal{I}_{\mathrm{CD}},~~~~\text{where} \non\\
\mathcal{I}_0 &=& \mathrm{Tr}\!\left[\mathcal{L}(\rho)H_0\right],~~~
\mathcal{I}_{\mathrm{CD}}
=
\mathrm{Tr}\!\left[\mathcal{L}(\rho)H_{\mathrm{CD}}\right],
\label{eq:heat1}
\ea
and we have used
$\mathrm{Tr}([\rho,H_{\mathrm{STA}}]H_{\mathrm{STA}})=0$.
Following Eq.~\eqref{eq:master} and Refs. \cite{mahunta2025shortcuts} and \cite{hartmann20multispin},
the bare heat current can be decomposed into independent momentum-mode contributions as
\begin{equation}
\mathcal{I}_0 = \sum_k \mathcal{I}_0^{(k)}, \quad \mathcal{I}_0^{(k)} = \mathrm{Tr}\!\left[\dot{\rho}_k H_{0k}\right] =\mathrm{Tr}\!\left[\mathcal{L}_k(\rho_k)H_{0k}\right],
\label{eq:heat_k_def}
\end{equation}

In contrast, the counterdiabatic contribution $\mathcal{I}_{\mathrm{CD}}$  vanishes   (see appendix \ref{App:heat_curr_power} for details), implying that the CD Hamiltonian does not
directly generate any heat current by itself. 
In the continuum thermodynamic limit, using Eqs.~(\ref{eq:spectrum}) and  (\ref{trajectory}), the total heat can be evaluated as 
\ba
\mathcal{I}_{\mathrm{STA}} &=& \mathcal{I}_0
=
\frac{1}{\pi}\int_0^{\pi}
\mathcal{I}_0^{(k)}\,dk ,\quad  \text{with} \non \\
\mathcal{I}_0^{(k)}
&=&
-\frac{\dot{\mu}\beta(\mu+\cos k)}
{1+\cosh(\beta E_k)}.
\label{eq:heat}
\ea

\textit{\textbf{(i) Low temperature regime: }} In low temperature limit of $\beta E_k >>1$ we get

\ba \label{eq:heat_low_temp} 
\mathcal{I}_{STA} \approx  -\frac{2 \dot{\mu}\beta}{\pi} \int_0^\pi (\mu+\cos(k)) e^{-\beta E_k} dk 
\ea 
Here the contribution from the modes with $E_k \gg 1/\beta$ are exponentially suppressed,
so that the dominant contribution to $\mathcal{I}_{STA}$ arises from the
low-energy critical modes only. This in turn results in the heat current being strongly reduced when the system is driven far from the critical region.

As shown in Fig. ~\ref{fig:Heat_power}(a), the strongly $\alpha$-dependent dispersion relation close to $\mu = -1$ result in  lower magnitudes of  heat currents for smaller $\alpha$ in this regime, thus once again showing the beneficial effects of LRIs in this regime. On the other hand, the opposite is true close to $\mu = +1$, where magnitudes of heat currents increase with increasing ranges of interactions  (see Appendix \ref{App:heat_curr_power}).

\textit{\textbf{(ii) High-temperature regime: }}
In the high-temperature limit of $\beta E_k\ll1$, 
we get an $\alpha$-independent form of $\mathcal{I}_{\mathrm{STA}}$  (see Eq. ~\eqref{eq:heat} and  Fig. ~\ref{fig:Heat_power}b):
\ba
\mathcal{I}_{\mathrm{STA}}
&\approx 
-\frac{1}{2\pi}\int_0^{\pi}
\dot{\mu}\beta(\mu+\cos k)\,dk =
-\frac{\dot{\mu}\beta\mu}{2}.
\label{eq:heat_highT}
\ea
 The absence of any critical features in this case can be attributed to strong thermal fluctuations dominating over quantum fluctuations
in the high-temperature regime, thereby effectively washing out any signature of the quantum criticality.

 \begin{figure*} 
    \centering
    \includegraphics[width=1\linewidth]{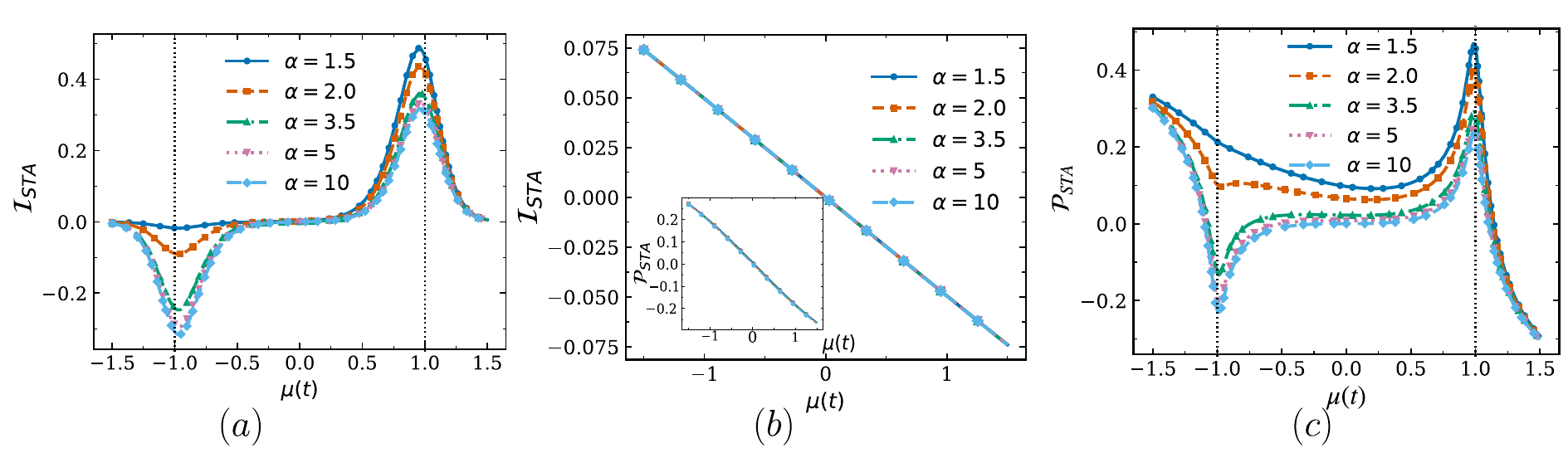}
    \caption{
(a) Heat current $\mathcal{I}_{\mathrm{STA}}$  in the low-temperature regime  $\beta=10$. 
(b) Heat current  $\mathcal{I}_{\mathrm{STA}}$ (main) and power $\mathcal{P}_{\mathrm{STA}}$ (inset) in the high-temperature regime $\beta=0.1$. 
(c) Power  $\mathcal{P}_{\mathrm{STA}}$ in the low-temperature regime $\beta=10$. 
}

    \label{fig:Heat_power}
\end{figure*}

\subsection{Power}
\label{secpow}

The total power generated during the
STA dynamics can be written as \cite{alicki79The}
\begin{align}
\mathcal{P}_{\mathrm{STA}}
&=
\mathrm{Tr}\!\left[\rho(t)\dot{H}_{\mathrm{STA}}\right]
=
\mathcal{P}_0+\mathcal{P}_{\mathrm{CD}},
\label{eqPsta}
\end{align}
where
\begin{align}
\mathcal{P}_0 &= \mathrm{Tr}\!\left[\rho(t)\dot{H}_0\right],
\qquad
\mathcal{P}_{\mathrm{CD}}
=
\mathrm{Tr}\!\left[\rho(t)\dot{H}_{\mathrm{CD}}\right].
\end{align}
the bare contribution to the power $\mathcal{P}_0$ can be decomposed into independent momentum-mode,
\begin{equation}
\mathcal{P}_0
=
\sum_k \mathcal{P}_0^{(k)},
\qquad
\mathcal{P}_0^{(k)}
=
\mathrm{Tr}\!\left[\rho_k \dot{H}_{0k}\right],
\label{eq:power_k_def}
\end{equation}

For a system evolving according to the master equation
[Eq.~\eqref{eq:master}], the $\mathcal{P}^{(k)}_{0}$ can be written as  $\mathcal{P}_0^{(k)} = \sum_n \lambda_n^{(k)} \dot{\tilde E}_n^{(k)},$
while
$\mathcal{P}_{\mathrm{CD}} = 0$  for an exact CD protocol (see appendix \ref{App:heat_curr_power} for detail).
In the continuum thermodynamic limit Eq. \eqref{eqPsta} we therefore get
\begin{equation}
\mathcal{P}_{\mathrm{STA}}
=
\mathcal{P}_0
=
\frac{1}{\pi}\int_0^{\pi}\mathcal{P}_0^{(k)}\,dk,
\end{equation}
with (see Eqs.~\eqref{eq:spectrum} and~\eqref{trajectory})
\begin{align}
\mathcal{P}_0^{(k)}
&=
\sum_n \lambda^{(k)}_n \dot{\tilde E}^{(k)}_n
=
-\lambda_1 \dot{\tilde E}_1^{(k)}
+\lambda_4 \dot{\tilde E}_4^{(k)} \nonumber\\
&=
-\frac{\dot{\mu}(\mu+\cos k)}{2E_k}
\tanh\!\left(\frac{\beta E_k}{2}\right).
\label{eqP0k}
\end{align}

\textit{\textbf{(i) Low-temperature regime.}}

In the low-temperature limit, $\beta E_k\gg 1$ we have
\begin{equation}
\mathcal P_{\mathrm{STA}}
\approx
-\frac{\dot\mu}{2\pi} \int_0^\pi \frac{\mu+\cos k}{E_k}\,dk,
\label{Eq:Total_Power_low_Temp}
\end{equation}
Here $E_k>0$ for all $k$. Consequently, the sign of $\mathcal P_{\mathrm{STA}}$ is determined by the $\left(\mu+\cos k\right)$ factors in Eq. \eqref{Eq:Total_Power_low_Temp}, weighted by the corresponding $E_k^{-1}$.
The different dispersion relations close to the two critical points result in critical signatures  in the form of local maxima in $|\mathcal P_{\mathrm{STA}}|$ to be strongly present only for short-range case ($ \alpha > 2$) close to $\mu = -1$, while the same remains present for all values of $\alpha$ close to $\mu = 1$.  In addition, $\mathcal P_{\mathrm{STA}}$ changes sign close to $\mu = -1$, while  $\mathcal P_{\mathrm{STA}}$ increases with increasing $\alpha$ close to $\mu = 1$ (see Fig.~\ref{fig:Heat_power}(c) and Appendix \ref{App:heat_curr_power}).  The absence of pronounced critical signatures for small $\alpha$ close to $\mu = -1$ can be attributed to $E_k$ vanishing slowly in this case in the long-range regime.

\textit{\textbf{(ii) High-temperature regime.}}
In the  limit of $\beta E_k\ll 1$ we have $\tanh(\beta E_k/2)\simeq \frac{\beta E_k}{2}$, which, analogous to the case of heat currents, result in an $\alpha$-independent form of $\mathcal{P}_{\mathrm{STA}}$ without any signature of criticality  (see Fig.~\ref{fig:Heat_power}(b))
\begin{align}
\mathcal{P}_{\mathrm{STA}}
&=
-\frac{\dot{\mu}\beta}{4\pi}\int_0^{\pi}(\mu+\cos k)\,dk
=
-\frac{\dot{\mu}\mu\beta}{4}.
\label{Eq:Total_Power_highT}
\end{align}

We summarize the results for power dissipation $\mathcal{P}_{\rm STA}$ and heat produced, $\mathcal{I}_{\rm STA}$ under STA protocol in Table \ref{tab:heat_power_summary}.

\begin{table*}[t]
\centering
\caption{Heat current $\mathcal{I}_{\rm STA}$ and power dissipated $\mathcal{P}_{\rm STA}$ near the critical points $\mu=\pm1$ in different temperature and interaction regimes.}
\renewcommand{\arraystretch}{1.5}

\begin{tabular}{|c|c|c|c|c|}
\hline
{\bf Observable} & {\bf $T$} & {\bf Criticality} &\multicolumn{2}{c|}{{\bf Variation with $\alpha$}} \\
\hline

\multirow{3}{*}[0pt]{$\mathcal{I}_{\rm STA}$}
& \multirow{2}{*}[0pt]{Low-$T$} 
& $\mu=-1$
& \multicolumn{2}{c|}{Suppressed for smaller $\alpha$}
\\ \cline{3-5}

& 
& $\mu=+1$
& \multicolumn{2}{c|}{Enhanced for smaller $\alpha$}
\\ \cline{2-5}

& High-$T$ 
& $\mu=\pm1$
& \multicolumn{2}{c|}{Does not depend on $\alpha$}
\\

\hline

\multirow{3}{*}[0pt]{$\mathcal{P}_{\rm STA}$}
& \multirow{2}{*}[0pt]{Low-$T$} 
& $\mu=-1$
& \multicolumn{2}{c|}{Shows local maxima in $|\mathcal{P}_{\rm STA}|$  for larger values of $\alpha$}
\\ \cline{3-5}

& 
& $\mu=+1$
& \multicolumn{2}{c|}{Enhanced for smaller $\alpha$}
\\ \cline{2-5}

& High-$T$ 
& $\mu=\pm1$
& \multicolumn{2}{c|}{Does not depend on $\alpha$}
\\

\hline
\end{tabular}

\label{tab:heat_power_summary}
\end{table*}

%%%%%%%%%%%%%%%%%%%%%%%%%%%%%%
\section{Charging a quantum battery through controlled dissipation}
\label{secbat}

Till now we have focussed on application of STA to drive a quantum critical system along its instantaneous thermal states. In contrast, now we show that the same technique, with some modifications, can be applied to charge a quantum battery \cite{rossini19quantum} through controlled dissipation, thereby emphasizing the versatile nature of the STA protocol studied here. In particular, we consider an evolution such that each mode is driven along the following trajectory:
\begin{equation}\label{athermal_rho}
\tilde\rho_k(t) = \sum_n \tilde{\lambda}^k_n(t) \ket{n_t^k}\bra{n_t^k}, \quad \text{with} \quad \tilde{\lambda}^k_n(t) = \frac{e^{\beta \tilde E^{(n)}_k}}{\tilde{\mathcal{Z}}_k},
\end{equation}
where $\tilde{\mathcal{Z}}_k = \sum_n e^{\beta \tilde E^{(n)}_k}$  is the normalization constant. 
Thus in contrast to the state considered in Eq. \eqref{trajectory}, here the modified target state Eq. \eqref{athermal_rho} is non-thermal, with higher energy states associated with higher occupation probabilities. As before,  the global state of the system is given by $\tilde \rho(t) = \bigotimes_k \tilde \rho_k(t).$

In order to realize an evolution constrained along the trajectory Eq.~(\ref{athermal_rho}), we consider the master equation (\ref{eq:master}). The counterdiabatic Hamiltonian $H_{\mathrm{CD}}$ Eq.(\ref{eq:Hcd}) guarantees transport along the instantaneous eigenstates $\ket{n^k_t}$, while the Lindblad jump operators $A^{(k)}_{mn}$  result in transitions between these instantaneous eigenstates; consequently these control terms remain unchanged under the current protocol. However, in contrast to the STA technique discussed above, now one needs to modify the rates $\gamma_{mn}^k$ to $\tilde\gamma_{mn}^{k} = \frac{\tilde{\lambda}_m}{r \tilde{\lambda}_n}$  in order to engineer a dissipation-induced quantum battery with non-zero ergotropy. Following the dispersion relation given in Eq. (\ref{eq:spectrum}), the rates of dispersion are modified as follows: $\tilde\gamma_{pn}^{(k)} = \gamma_{qn}^{(k)}$,  $\tilde\gamma_{np}^{(k)} = \gamma_{nq}^{(k)}$ for $p, q = 1,4,~p \neq q$ and $n = 2,3$, $\tilde\gamma_{mn}^{(k)} = \gamma_{nm}^{(k)}$ for $m, n = 1,4;~m\neq n$, and  $\tilde\gamma_{23}^{(k)} = \tilde\gamma_{32}^{(k)} = \gamma_{23}^{(k)} = \gamma_{32}^{(k)}$.

\subsection{Ergotropy}
Ergotropy is the maximum amount of work that can be extracted from a quantum state $\rho(t)$ by a unitary operation without altering the von Neumann entropy of the state. 
In the present case the instantaneous total ergotropy is given by \cite{campaioli17enhancing,bhattacharjee2021quantum,allahverdyan2004maximal,alicki13entanglement}
\ba
\mathcal{W}_{\text{tot}}(t) &=& \frac{1}{\pi}\int_0^{\pi} \mathcal{W}_k(t) dk \non\\
\mathcal{W}_k(t) &=& \mathrm{Tr}\left[\tilde\rho_k(t) H_{0k}(t)\right] - \mathrm{Tr}\left[\Pi_k(t) H_{0k}(t)\right].
\label{eq:ergotropy_k}
\ea
Here $\Pi_k(t) = \mathcal{U}(t) \tilde{\rho}_{k}(t) \mathcal{U}^{\dagger}(t)$ with  $\mathcal{U}(t)  = e^{i\phi(t)} \left( \ket{\xi^k_1(t)} \bra{\xi^k_4(t)} + \ket{\xi^k_4(t)}\bra{\xi^k_1(t)}\right)$ (see Eq. \eqref{levelsk})
denotes the passive state corresponding to the population-inverted state $\tilde{\rho}_{k}(t)$, for an arbitrary phase factor $\phi(t)$. In the present case $\Pi_k(t)$ coincides with the instantaneous thermal state $\rho_k^{th}(t)$ at time $t$ \cite{pusz1978passive,lenard1978thermodynamical}.  
Using this construction together with the spectral decomposition of the Hamiltonian, the ergotropy of each mode can be written as
\begin{eqnarray} \label{eq:ergotropy_total}
    \mathcal{W}_k(t) &=& \sum_n \tilde E_{k}^{(n)} ( \tilde{\lambda}^{(k)}_n(t) -\lambda^{(k)}_n(t)) \non\\
    &=& 2 E_k(\alpha) \tanh{\left(\beta E_k(\alpha)/2\right)}
\end{eqnarray}
The dependence of the ergotropy on $\alpha$  for different values of $\mu$ is shown in Fig.~\ref{fig:ergotropy}(a). The relation $E_k \propto k^{\alpha-1}$ for $1 \leq \alpha \leq 2$ close to $\mu = -1$ results in long-range advantage in charging a quantum battery in this regime, signified by increasing values of ergotropy for decreasing $\alpha$. On the other hand, the linear dispersion relation of $E_k$ for all $\alpha$ results in an ergotropy which is weakly dependent on $\alpha$ for $\mu \approx 1$.

As shown in Fig.~\ref{fig:ergotropy}(b), lower temperatures correspond to higher ergotropy for all $\alpha$, thus
highlighting the importance of low temperatures for charging a quantum battery. On the other hand, thermal mixing at high temperatures suppresses
population imbalance and reduces the extractable work.

Interestingly, despite the system being driven across critical points, the ergotropy exhibits no signatures of
criticality. This may be attributed to the control protocol studied here, which ensures that the system is  evolved through a particular trajectory, irrespective of the presence of criticality.

\begin{figure}
    \centering
    \includegraphics[width=01\linewidth]{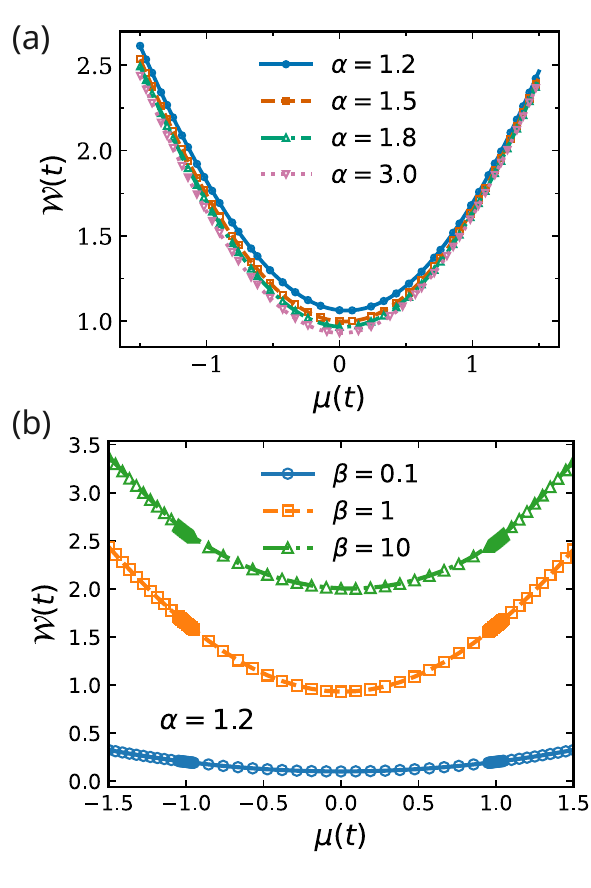}
    \caption{(a) Ergotropy $\mathcal{W}(t)$  with $\mu$ for different values of  $\alpha$ at inverse temperature $\beta =1$. The ergotropy increases with increasing range of interactions.  (b) Variation of the ergotropy $\mathcal{W}(t)$  with $\mu$ for different values of $\beta$, for $\alpha = 1.2$. Lower temperatures result in higher values of ergotropy.}
    \label{fig:ergotropy}
\end{figure}

\subsection{Power and Heat current} 
\label{secbatteryPI}
Following Sec. (\ref{secPower_Heat}) and Ref.-\cite{mahunta2025shortcuts}, the  power for the mode $k$ can be written as  $\mathcal{\tilde{P}}_{0}^{(k)} = \sum_n \tilde\lambda_n^k \dot{E}^k_n = \dot{E}_k \left( \tilde{\lambda_1} - \tilde\lambda_4 \right) = -\mathcal{{P}}_{0}^{(k)}$, while the corresponding heat current is given by  $\mathcal{\tilde{I}}_{0}^{(k)} = \sum_n \dot{\tilde\lambda}_n^k E^k_n = E_k \left( \dot{\tilde\lambda}_1 - \dot{\tilde\lambda}_4 \right) = -\mathcal{{I}}_{0}^{(k)}$. As before, the modified CD Hamiltonian does not directly contribute to the power or the heat current in this case as well. Consequently, we finally get
\ba
\mathcal{\tilde{P}}_{STA} &=& \frac{1}{\pi}\int_0^\pi \mathcal{\tilde{P}}_{0}^{(k)} dk = -\mathcal{P}_{\text{STA}} \non\\
\mathcal{\tilde{I}}_{STA} &=& \frac{1}{\pi}\int_0^\pi \mathcal{\tilde{I}}_{0}^{(k)}  dk = -\mathcal{I}_{\text{STA}}.
\ea

\section{Conclusions}
\label{sec:conclusion}
We studied shortcuts to adiabaticity (STA) in an open long-range quantum critical system, through suitable engineered unitary and non-unitary operators. We show that for suitable choices of parameters, LRIs can be significantly beneficial for controlling excitations in open quantum critical systems driven out of equilibrium, as well as for charging a quantum battery through a STA inspired quantum control technique. 

In this work we have considered the LRK chain, which shows two different dispersion relations close to the two quantum critical points $\mu = \pm 1$;
the energy gap $E_k(\alpha)$ increases with increasing range of interactions for $\mu = -1$, which in turn results in the long range advantages discussed above. In particular, for LRIs quantified by $1 <  \alpha < 2$, the CD Hamiltonian strength $h_m$ denoting the strength of interactions between particles separated by a distance $m$ decays  algebraically with $m$ both away from criticality, as well as at the criticality. This is in sharp contrast with the behavior seen for short range interactions obtained for $\alpha > 2$, in which case $h_m$ assumes a $m$-independent constant value at criticality, analogous to that seen earlier for the transverse Ising chain in Ref. \cite{campo12assisted}. This result provides a clear way of harnessing LRIs for engineering practically realizable exact STA protocols in quantum critical systems, which otherwise can be expected to be impossible to implement, owing to the requirement of infinite-range interactions at criticality. 
In case of control in the presence of dissipation, LRIs assist in STA through reduction in the costs  of control, quantified by decrease of magnitude of heat current with decreasing values of $\alpha$. Finally we propose a modified STA protocol, which aims at using the setup as a quantum battery, through population inversion. We show that longer ranges of interactions are beneficial in this case as well, through enhancement of ergotropy for lower values of $\alpha$, for $\mu \approx -1$.

In sharp contrast to the beneficial effects of LRIs discussed above, range of interactions have distinctly different effects close to $\mu = 1$. Here $E_k(\alpha)$ assumes linear dispersion relation for all $\alpha > 1$, with $E_k(\alpha)$ decreasing  with decreasing $\alpha$. This in turn results in vanishing of the long-range advantages observed earlier for $\mu \to -1$.

The above discussed effects of LRIs and criticality become more pronounced for lower values of temperatures. On the other hand, thermal fluctuations dominate at high temperatures, thereby washing away all effects of $\alpha$ and criticality. Further, as discussed in Appendix \ref{Appreal}, inspite of the local form of the control Lindblad operators in the momentum space, the real space representation of the same involves multi-body interaction terms even away from criticality, for all values of $\alpha$. This can be attributed to the requirement for control of entropy, which in general can be expected to involve non-local control terms.

Finally, we note that the STA protocol presented in this work can be expected to be implementable in currently existing experimental platforms, including in ion traps \cite{rossnagel16a} and quantum simulators \cite{bernien17probing, King22}, and  can be highly relevant for quantum control in many-body open quantum systems \cite{claeys19floquet}, as well as for applications in quantum technologies \cite{mukherjee24promises}, including finite-time quantum engines \cite{cangemi24quantum} and quantum batteries \cite{campaioli2019quantum}. 

\section*{ACKNOWLEDGEMENTS}
VM acknowledges A. del Campo for fruitful discussions on related works. VM also acknowledges support from Anusandhan National Research Foundation (ANRF) through ARG (Project No.
ANRF/ARG/2025/002531/PS).

\appendix
%%--------------------- Appendix A --------------
\section{Low-momentum expansion of the pairing function and quasiparticle spectrum}
\label{App:A}

In this appendix, we derive the low-momentum behavior of the  function $f_\alpha(k)$ [Eq.~\eqref{eq:f_alpha}] and the quasiparticle spectrum $E_k$ [Eq.~\eqref{eq:Ek_minus_one}] near the critical modes $k=0$ and $k=\pi$.

\subsection{Expansion around \texorpdfstring{$\mu=-1$}{mu=-1}}

Near the critical point $\mu=-1$, where the gap closes at $k=0$, we have~\cite{vodola2016long,dutta2017probing,olver2010nist,abramowitz1964handbook}
\begin{equation}
f_\alpha(k)\sim
\begin{cases}
A_\alpha k^{\alpha-1}, & 1<\alpha<2,\\[4pt]
k\ln(1/k), & \alpha=2,\\[4pt]
B_\alpha k, & \alpha>2,
\end{cases}
\label{eq:falpha_smallk}
\end{equation}
where
\[
A_\alpha=
\left|\Gamma(1-\alpha)\cos\!\left(\frac{\pi\alpha}{2}\right)\right|,
\qquad
B_\alpha=\zeta(\alpha-1).
\]
Defining $\epsilon_-=\mu+1$ and expanding $\mu+\cos k\simeq \epsilon_--k^2/2$, the quasiparticle spectrum reduces to (see Fig. \ref{fig:energy_alpha})
\begin{equation}
E_k^2\sim
\begin{cases}
\epsilon_-^2+A_\alpha^2k^{2(\alpha-1)} & 1<\alpha<2,\\[4pt]
\epsilon_-^2+k^2\ln^2(1/k), & \alpha=2\\[4pt]
\epsilon_-^2+B_\alpha^2k^2, & \alpha>2
\end{cases}
\label{eq:Ek_scaling_minus}
\end{equation}

\begin{figure}
    \centering
    \includegraphics[width=1\linewidth]{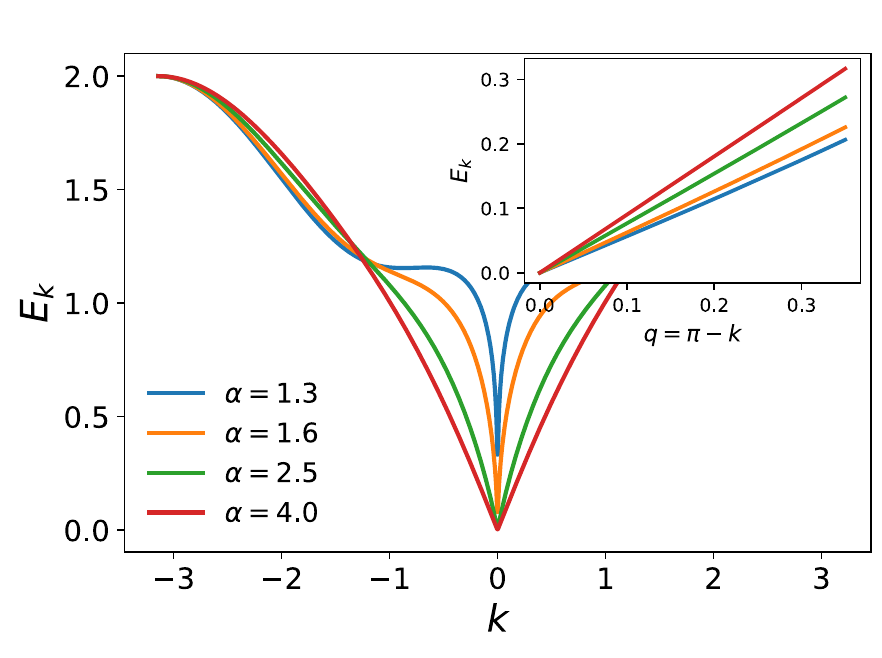}
    \caption{Quasiparticle spectrum $E_k$ of the LRK model for different   $\alpha$  at $\mu =-1 $. The inset shows the dispersion near $k=\pi$, and at $\mu =+1$ where the spectrum remains linear for all $\alpha$.}
    \label{fig:energy_alpha}
\end{figure}

\subsection{Expansion around \texorpdfstring{$\mu=+1$}{mu=+1}}

Near the second critical point $\mu=+1$, the gap closes at $k=\pi$. Writing $q=\pi-k$, the pairing function expands as~\cite{solfanelli2023quantum}
\begin{equation}
f_\alpha(\pi-q)\sim \tilde B_\alpha q,
\qquad
\tilde B_\alpha=\eta(\alpha-1),
\label{eq:falpha_smallpi}
\end{equation}
where $\eta(s)=(1-2^{1-s})\zeta(s)$ is the Dirichlet eta function. Defining $\epsilon_+=\mu-1$ and expanding $\mu+\cos(\pi-q)\simeq \epsilon_++q^2/2$, the spectrum becomes
\begin{equation}
E_k^2\sim \epsilon_+^2+\tilde B_\alpha^2q^2 .
\label{eq:Ek_scaling_plus}
\end{equation}

Hence, unlike the case $\mu=-1$, the quasiparticle spectrum at $\mu=+1$ remains linear for all $\alpha>1$.
%%------------End of the Appendix_A------------------

\section{CD  drive for the LRK chain}
\label{App:CD_LRK}

In this section we briefly derive the CD Hamiltonian
for the LRK chain in Fermionic representation, and its corresponding real space representation
\subsection{Momentum-space representation}
\label{AppHCD}
The momentum-space representation of the 
long-range Kitaev chain (Eq.\ref{bare_hamilotonian}) reads
\begin{equation}
H_0(t)=
\sum_k
\Psi_k^\dagger
\mathcal H_k(\mu)
\Psi_k ,
\qquad
\Psi_k^\dagger=(c_k^\dagger,c_{-k}),
\end{equation}
with $\mathcal H_k(\mu) = (\mu+\cos k)\sigma^z + f_\alpha(k)\sigma^x$ and the quasiparticle spectrum is $ E_k(\mu,\alpha) = \sqrt{(\mu+\cos k)^2+f_\alpha^2(k)} .$ Each momentum sector therefore corresponds to an effective
two-level Hamiltonian
\(
\mathcal H_k=\vec d_k\cdot\vec\sigma
\)
with $
\vec d_k=
\left(
f_\alpha(k),\,0,\,\mu+\cos k
\right).$
 For a generic two-level Hamiltonian the adiabatic gauge potential
is given by \cite{campo12assisted, kolodrubetz17geometry}
\begin{equation}
A_\mu^{(k)}
=
\frac12
\frac{\vec d_k\times\partial_\mu\vec d_k}{|\vec d_k|^2}
\cdot\vec\sigma .
\end{equation}

Since
\(
\partial_\mu(\mu+\cos k)=1
\)
and
\(
\partial_\mu f_\alpha(k)=0
\),
we obtain
\begin{equation}
\vec d_k\times\partial_\mu\vec d_k
=
(f_\alpha,0,\mu+\cos k)\times(0,0,1)
=
(0,-f_\alpha,0).
\end{equation}

Thus the adiabatic gauge potential becomes
\begin{equation}
A_\mu^{(k)}
=
-\frac12
\frac{f_\alpha(k)}{E_k^2}
\sigma^y, 
\end{equation}
which results in the counterdiabatic Hamiltonian Eq.~(\ref{eq:HCD_momentum}):
\ba
H_{\rm CD}(t)
&=&
\dot\mu(t)
\sum_k A_\mu^{(k)}\non\\
&=&
-\frac{\dot\mu(t)}{2}
\sum_k
\frac{f_\alpha(k)}{E_k^2}
\Psi_k^\dagger\sigma^y\Psi_k .
\label{AppeqHCD}
\ea

\subsection{Real-space representation of $H_{CD}$}
\label{appreal}
To obtain the real-space form of the CD Hamiltonian we first note the
identity
\begin{equation}
\Psi_k^\dagger\sigma^y\Psi_k
=
i\left(
c_{-k}c_k
-
c_k^\dagger c_{-k}^\dagger
\right).
\end{equation}
Substituting this into Eq.~(\ref{AppeqHCD}) gives
\begin{equation} \label{appeq:cd_k}
H_{\rm CD}(t)
=
-\frac{\dot\mu(t)}{2}
\sum_k
\frac{f_\alpha(k)}
{(\mu+\cos k)^2+f_\alpha^2(k)}
\,i\left(
c_{-k}c_k
-
c_k^\dagger c_{-k}^\dagger
\right).
\end{equation}
We now express the Fermionic operators in real space using the inverse
Fourier transform
\begin{equation}
c_k=\frac{1}{\sqrt N}\sum_j e^{-ikj}c_j,
\qquad
c_{-k}=\frac{1}{\sqrt N}\sum_l e^{ikl}c_l .
\end{equation}
This yields
\begin{equation}
c_{-k}c_k
=
\frac{1}{N}
\sum_{j,l}
e^{ik(l-j)}
c_l c_j .
\end{equation}
Introducing the separation $m=j-l$ and reorganizing the sums, the above
expression can be written as
\begin{equation}
c_{-k}c_k
=
\frac{1}{N}
\sum_{j}\sum_{m}
e^{-ikm}\,
c_j c_{j+m}.
\end{equation}
Substituting this result back into the CD Hamiltonian(Eq.~\ref{appeq:cd_k}) and performing the
summation over momentum, one obtains Eq. \eqref{eq:Cd_qcp}.

%%%%%%%%%%%%%%%%%%%%%%%% 
%---------------Derivation -----
\subsection{Derivation for the CD coupling \texorpdfstring{$h_m$}{h\_m}} 
Here, we derive the form of CD coupling $h_m(\mu, \alpha)$ across the criticality $\mu =\pm1$, discussed in
Sec.~\ref{Unitary_CD_drive_QCP}. 

% \subsubsection*{A. Critical point \texorpdfstring{$\mu=-1$}{mu=-1} in the long-range regime \texorpdfstring{$1<\alpha<2$}{1<alpha<2}}
(a) $\bm{\mu = -1,~1<\alpha < 2:}$ Putting $\epsilon_- =0 $ in Eq. \eqref{eq:h_r_LR} one gets
\begin{equation*}
    h_m(\epsilon_-, 1<\alpha < 2) \approx \frac{1}{4\pi A_\alpha}\int_0^\pi  k^{(1- \alpha)}\sin(mk) \, dk 
\end{equation*}
%%%%%%%%%%%%%%%%%%%%%%%%%%%%%%%%%
Making the scaling substitution $q=mk$ we get
\begin{equation*}
h_m \approx
\frac{m^{\alpha-2}}{4\pi A_\alpha}
\int_0^{m\pi} q^{1-\alpha}\sin q\, dq.
\end{equation*}
The above integral converges for large $m$ and using  the standard identity \cite{Gradshteyn}
\begin{equation}
\int_0^\infty q^{\beta-1}\sin q\, dq
=
\Gamma(\beta)\sin\!\left(\frac{\pi\beta}{2}\right),
\qquad 0<\beta<1,
\label{eq:sin_integral}
\end{equation}
with $\beta=2-\alpha$, we obtain
\begin{equation}
h_m
\approx
\frac{m^{\alpha-2}}{4\pi A_\alpha}
\Gamma(2-\alpha)\sin\!\left(\frac{\pi\alpha}{2}\right).
\end{equation}
Using $\Gamma(2-\alpha)=(1-\alpha)\Gamma(1-\alpha)$,
one finally obtains Eq. \eqref{eq:hm_LR_critical} in the main text.

%%%%%%%%%%%%%%%%%%%%%%%%%%%%%%%%%%%%%%---------------
%%%%%%%%%%%%%---------------
(b) $\bm{\mu \to -1, \alpha > 2:}$
The integral (Eq.\eqref{hmalgtr2}) 
\begin{equation}
h_m(\epsilon_-,\alpha>2)
\approx
\frac{1}{4\pi}
\int_0^\pi
\frac{B_\alpha k\,\sin(mk)}
{\epsilon^2+B_\alpha^2 k^2}\, dk 
\end{equation}
is dominated by the  momentum region
near $k=0$. Therefore the upper limit of integration can be extended to infinity.
Factoring out $B_\alpha^2$ from the denominator gives
\begin{equation}
h_m(\epsilon_-,\alpha>2)
\approx
\frac{1}{4\pi B_\alpha}
\int_0^\infty
\frac{k\,\sin(mk)}
{k^2+\kappa^2}\, dk,
\quad
\kappa=\frac{|\epsilon|}{B_\alpha}.
\end{equation}
Using the standard integral \cite{Gradshteyn}
\begin{equation}
\int_0^\infty
\frac{k\sin(rk)}{k^2+\kappa^2}\, dk
=
\frac{\pi}{2}e^{-\kappa r},
\label{eq:ref_integration}
\end{equation}
we obtain Eq. \eqref{eq:hr_SR_minus}.

(c) $\bm{\mu \to +1, \alpha:}$ Now, we derive the CD coupling $h_m(\mu, \alpha) $ near second critical point $\mu\to +1$, where the gap closes
at $k=\pi$. Small $q=\pi -k$ expansions of pairing function and quasi particle spectrum are given in Eqs. \eqref{eq:falpha_smallpi}-\eqref{eq:Ek_scaling_plus} and using them in Eq.\eqref{eq:Cd_qcp}, one finds

\begin{equation}
h_m(\epsilon_+,\alpha)
\simeq
\frac{(-1)^{m+1}}{4\pi}
\int_0^\pi dq\;
\frac{\widetilde B_\alpha q\sin(mq)}
{\epsilon_+^2+\widetilde B_\alpha^2 q^2}.
\label{App:eq:hm_plus_start}
\end{equation}
Where we have used the sine factor as, $\sin(mk)=\sin[m(\pi-q)]=(-1)^{m+1}\sin(mq)$. 
Proceeding exactly as in the $\mu=-1$, $\alpha>2$ case and  defining: 
$ \kappa_+=\frac{|\epsilon_+|}{\widetilde B_\alpha}.$ we get
\begin{equation}
h_m(\epsilon_+,\alpha)
\simeq
\frac{(-1)^{m+1}}{4\pi\widetilde B_\alpha}
\int_0^\infty dq\;
\frac{q\sin(mq)}
{q^2+\kappa_+^2}.
\end{equation}
Using the same standard integral(see Eq.\eqref{eq:ref_integration} we get
\begin{equation}
h_m(\epsilon_+,\alpha)
\simeq
(-1)^{m+1}
\frac{1}{8\widetilde B_\alpha}
\exp\left[-\frac{|\epsilon_+|}{\widetilde B_\alpha}m\right].
\end{equation}
Finally, substituting $\widetilde B_\alpha=\eta(\alpha-1)$ gives us Eq.~\eqref{eq:hr_SR_plus}.

\section{Lindblad operators and dissipation rates}
\label{app:Lindblad_operators_and_rates}
In this appendix, we provide the engineered Lindblad operators, $A_{mn}^{(k)}$ and transition rates $\gamma^{(k)}_{mn}$. Without loss of generality, we focus on the one of operators, $A_{12}^{(k)}$ as follows, 
\ba
&&A_{12}^{(k)}(t) = \ket{\xi^k_1}\bra{\xi^k_2} \non\\ 
&=& \frac{\phi^{(k)}_t \ket{1_k, 1_{-k}}\bra{1_k, 0_{-k}} + \ket{0_k, 0_{-k}}\bra{1_k, 0_{-k}}}{\sqrt{\left|\phi_t^{(k)}\right|^2 + 1}}.
\label{eq:L12appB}
\ea 
Following \cite{mahunta2025shortcuts} we have
\ba
\ket{1_k, 1_{-k}}\bra{1_k, 0_{-k}} &=& c_k^{\dagger}c_{-k}^{\dagger}\ket{0_k, 0_{-k}}\bra{0_k, 0_{-k}}c_k \non\\&=& c_k^{\dagger}c_{-k}^{\dagger} c_k,
\label{eqL12_1appB}
\ea
Similarly, 
\ba
&&\ket{0_k, 0_{-k}}\bra{1_k, 0_{-k}} = c_k c_{-k}  c_{-k}^{\dagger}.
\label{eqL12_2appB}
\ea
Following Eqs. \eqref{eq:L12appB} - \eqref{eqL12_2appB} we get
\ba
A_{12}^{(k)}(t) = \frac{\phi^{(k)}_t c_k^{\dagger}c_{-k}^{\dagger} c_k + c_k c_{-k}  c_{-k}^{\dagger}}{\sqrt{|\phi_t^{(k)}|^2 + 1}}.
\label{eqL12_3appB}
\ea 

Proceeding as above, one can show that the different Lindblad operators introduced in Eqs. \eqref{eq:gamma} and \eqref{levelsk} are given by:
\begin{align}
A_{12}^{(k)}(t) &= \left(A_{21}^{(k)}\right)^{\dagger} = \frac{\phi^{(k)}_t c_k^{\dagger}c_{-k}^{\dagger} c_k + c_k c_{-k}  c_{-k}^{\dagger}}{\sqrt{\left|\phi_t^{(k)}\right|^2 + 1}}, \nonumber \\
A_{13}^{(k)} &= \left(A_{31}^{(k)}\right)^{\dagger} =  \frac{\phi_t^{(k)} c_k^{\dagger} c_{-k}^{\dagger} c_{-k} + c_k c_{-k} c_k^{\dagger}}{\sqrt{\left(\phi_t^{(k)}\right)^2 + 1}}, \nonumber \\
\begin{split}
A_{14}^{(k)} &= \left(A_{41}^{(k)}\right)^{\dagger} \\
&= \frac{-\phi_t \theta_t c_k^{\dagger} c_{-k}^{\dagger} c_k c_{-k} + \phi_t c_k^{\dagger} c_{-k}^{\dagger} - \theta_t c_k c_{-k}  +  c_k c_{-k} c_k^{\dagger} c_{-k}^{\dagger}}{\sqrt{\left(\left(\phi_t\right)^2 + 1\right) \left(\left(\theta_t\right)^2 + 1 \right)}}, 
\end{split} \nonumber \\
A_{23}^{(k)} &= \left(A_{32}^{(k)}\right)^{\dagger} = c_k^{\dagger} c_{-k}, \nonumber \\
A_{24}^{(k)} &= \left(A_{42}^{(k)}\right)^{\dagger} = \frac{-\theta_t^{(k)} c_k^{\dagger} c_k c_{-k} + c_{-k} c_k^{\dagger} c_{-k}^{\dagger}}{{\sqrt{\left(\theta_t^{(k)}\right)^2 + 1}}}, \nonumber \\
A_{34}^{(k)} &= \left(A_{43}^{(k)}\right)^{\dagger} = \frac{-\theta_t^{(k)} c_{-k}^{\dagger} c_k c_{-k} + c_{k} c_k^{\dagger} c_{-k}^{\dagger}}{{\sqrt{\left(\theta_t^{(k)}\right)^2 + 1}}}.
\label{Lks}
\end{align}

\begin{widetext}

The \(\gamma_{mn}^{(k)}\)s for the mode \(k\) are given by 
\ba
\gamma_{12}^{(k)} &=& \gamma_{13}^{(k)}  = \frac{\beta e^{\beta E_k}  \left(\mu + \cos k \right) \dot{\mu}}{2\left(1 + e^{\beta E_k} \right)E_k},\hspace{3cm}\quad
\gamma_{21}^{(k)} = -\frac{\dot{\mu} \beta e^{-\beta E_k } \left(\mu + \cos k \right) \tanh \left[\beta E_k/2 \right]}{4 E_k},\non\\
\gamma_{23}^{(k)} &=& \gamma_{32}^{(k)}=-\frac{\beta \dot{\mu} \left(\mu + \cos k \right) \tanh \left[\beta E_k / 2 \right]}{4 E_k},\hspace{1.5cm} \quad
\gamma_{31}^{(k)} = -\frac{\dot{\mu}\beta e^{-\beta E_k} \left(\mu + \cos k \right) \tanh\left[\beta E_k / 2 \right]}{4 E_k}, \non\\
\gamma_{34}^{(k)} &=& \gamma_{24}^{(k)} =  -\frac{\dot{\mu}\beta e^{\beta E_k}\left(\mu + \cos k \right) \tanh \left[\beta E_k/2 \right]}{4 E_k}, \hspace{0.89cm}\quad
\gamma_{43}^{(k)} = \gamma_{42}^{(k)} = -\frac{ \dot{\mu}\beta \left( \mu + \cos k\right)}{2\left(1 + e^{\beta E_k} \right)E_k},\non\\
\gamma_{14}^{(k)} &=& \frac{\dot{\mu}\beta e^{2\beta E_k} \left(\mu + \cos k \right)}{2 \left(1 + e^{\beta E_k} \right)E_k}, \hspace{4cm}\quad
\gamma_{41}^{(k)} = -\frac{\dot{\mu}\beta e^{-\beta E_k} \left(\mu + \cos k \right)}{2\left(1 + e^{\beta E_k} \right)E_k}.
\label{eq:gamk}
\ea

\end{widetext}

In low temperature limit, \(\beta E_k \gg 1\) the \(\gamma_{mn}^{(k)}\) reduces to;

\ba \gamma_{12}^{(k)} &=& \gamma_{13}^{(k)} \approx \frac{\dot{\mu}\beta \left(\mu + \cos k \right)}{2 E_k} \non \\
\gamma_{23}^{(k)} &=& \gamma_{32}^{(k)} = -\frac{ \dot{\mu} \beta \left(\mu + \cos k \right)}{4 E_k} \non
\ea

\ba
\gamma_{34}^{(k)} &=& \gamma_{24}^{(k)} = -\frac{\dot{\mu} \beta e^{\beta E_k}\left(\mu + \cos k \right)}{4 E_k}\non\\ 
\gamma_{14}^{(k)} &=& \frac{\dot{\mu}\beta e^{\beta E_k} \left(\mu + \cos k \right)}{2 E_k}\non\\ 
\gamma_{41}^{(k)} &=& \gamma_{42}^{(k)}, \gamma_{43}^{(k)}, \gamma_{31}^{(k)}, \gamma_{21}^{(k)} \approx  0. \label{eq:gamk_low_temp}
\ea
In the high-temperature regime $\beta E_k \ll 1$ we have  $\tanh(\beta E_k/2)\approx \beta E_k/2$. Consequently, in this case the engineered transition rates $\gamma_{mn}^{(k)}$ that scale as $
\tanh(\beta E_k/2)/E_k $, 
such as $\gamma_{21}$ and $\gamma_{23}$, become independent of the energy $E_k$, and hence of the long-range exponent $\alpha$ (see Fig.~\ref{fig:gamma_low_highTemp}f).

%%%%%%%%%%%%%%%%%%%%%%%%%%%%%%%%%%%%%%%%%%%%%%%%%%%%%%%%%

\begin{figure*}
    \centering
    \includegraphics[width=\linewidth]{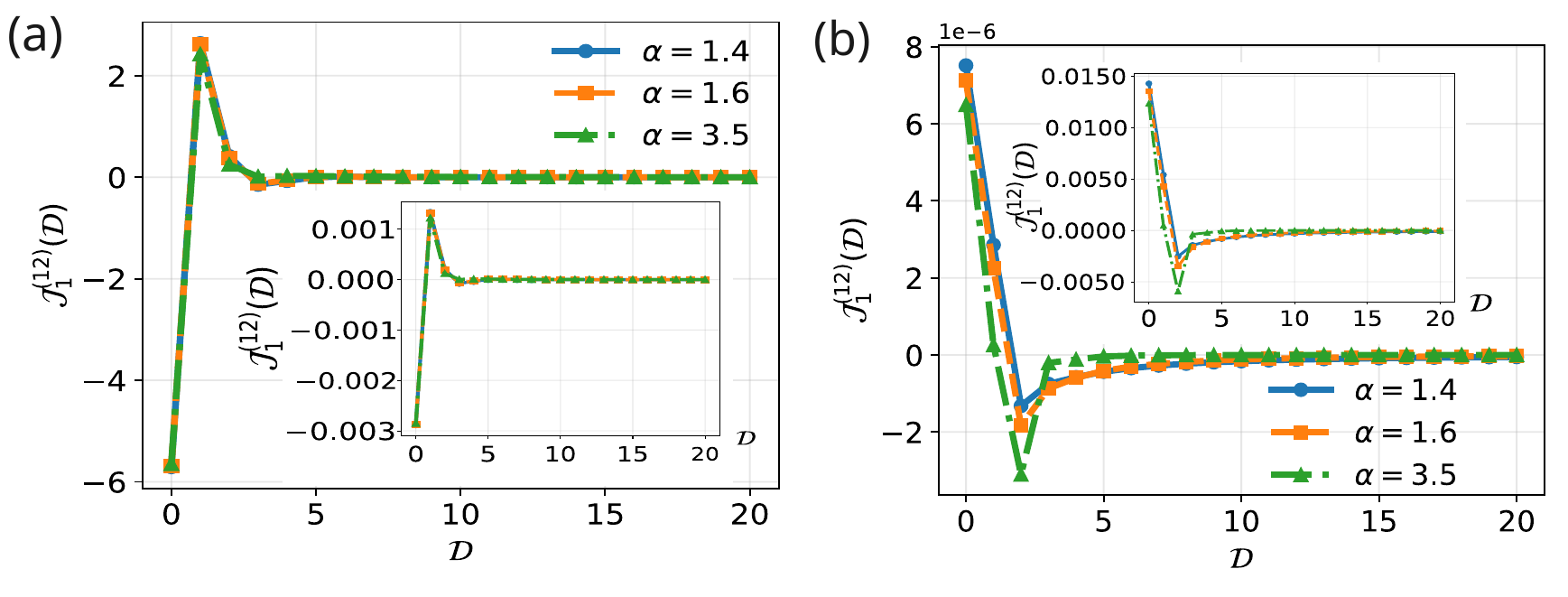}
    \caption{ Variation of the non-unitary control strength $\mathcal{J}_{12}$ with $\mathcal{D}$ for different values of the long-range exponent $\alpha$ in the vicinity of the critical points (a) $\mu=-1$ and (b) $\mu=+1$, in the low-temperature regime ($\beta=10$). Insets show the corresponding behaviors in the high-temperature regime ($\beta=0.01$).}
    \label{fig:J12_with_D}
\end{figure*}

%%%%%%%%%%%%%%%%%%%%%%%%%%%%%%%%%%%%%%%%%%%%%%
\section{Heat currents and Power }
\label{App:heat_curr_power}
\textbf{Heat Current:} 
As defined in the Eq.\eqref{eq:heat1}, the CD contribution to heat dissipation is given as, 
\ba 
\mathcal{I}_{\rm CD} = \mathrm{Tr}\!\left[\mathcal{L}(\rho)H_{\rm CD}\right];~~  \mathcal{L}(\rho) = \sum_m \dot\lambda_m \ket{m_t} \bra{m_t}
\ea
Using the CD Hamiltonian $H_{\rm CD}$(C.f Eq.\eqref{eq:Hcd}) one gets,  
\begin{equation*}
\begin{aligned}
\mathcal{I}_{CD}
=
i\,\mathrm{Tr}\Bigg[
\sum_m \dot{\lambda}_m |m_t\rangle \langle m_t|
\sum_n
\Big(
|\partial_t n_t\rangle \langle n_t|
\Big)
\Bigg]
\\
-i\,\mathrm{Tr}\Bigg[
\sum_m \dot{\lambda}_m |m_t\rangle \langle m_t|
\sum_n
\Big(
\langle n_t|\partial_t n_t\rangle
|n_t\rangle \langle n_t|
\Big)
\Bigg].
\end{aligned}
\end{equation*}
 
One can further simplify $\mathcal{I}_{\rm CD}$ as \cite{mahunta2025shortcuts}, 
\begin{align*}
\mathcal{I}_{CD}
&=
i\,\mathrm{Tr}\!\left[
\sum_{m,n}
\dot{\lambda}_m
|m_t\rangle
\langle m_t|\partial_t n_t\rangle
\langle n_t|
\right]
\nonumber\\[0.2cm]
&\quad
-i\,\mathrm{Tr}\!\left[
\sum_{m,n}
\dot{\lambda}_m
\langle n_t|\partial_t n_t\rangle
|m_t\rangle
\langle n_t|
\right]
\nonumber\\[0.2cm]
&=
\sum_m \dot{\lambda}_m \Phi_{mm}
-
\sum_l \dot{\lambda}_l \Phi_{ll}
=0.
\end{align*}
where we have used, $\Phi_{mm} = \langle m_t | \partial_t m_t \rangle $. \\

\textbf{Power:} Next, we evaluate the power dissipated $\mathcal{P}_{\rm CD} = \mathrm{Tr} [ \rho(t) \dot{H}_\mathrm{CD}(t)]$ due to the CD Hamiltonian $H_{\rm CD}$  in presence of a non-unitary evolution by considering  the desired  path for the system to be the instantaneous thermal state  $\rho(t) = \sum_l \lambda_l(t)\ket{l_t}\bra{l_t}$. One can use the completeness relation,  $\sum_l | l_t \rangle \langle l_t | = 1$ to get
\begin{align}
   &\partial_t^2 \left( \sum_l | l \rangle \langle l | \right ) = \sum_l | \ddot{l} \rangle \langle l | + 2 | \dot{l} \rangle \langle \dot{l} | + | l \rangle \langle \ddot{l} | = 0 \nonumber \\
   &\Rightarrow \sum_l | \dot{l} \rangle \langle \dot{l} | =
 -\dfrac{1}{2} \sum_l \left(| l \rangle \langle \ddot{l} | + | \ddot{l} \rangle \langle l | \right).
\label{eq_trick1}
\end{align}
Further, the normalization condition $ \langle l | l \rangle = 1$ leads to 
\begin{align}
	 \langle \dot{l} | l \rangle = - \langle l | \dot{l} \rangle~~ \Rightarrow~~ 2 \langle l | \dot{l} \rangle  = \langle l | \dot{l} \rangle - \langle \dot{l} | l \rangle.
      \label{eq_normalisation}
  \end{align}
Now we use Eq.(\ref{eq_trick1}) and Eq.(\ref{eq_normalisation}) to evaluate the time derivative of the  $H_\mathrm{CD}(t)$ (C.f Eq.(\ref{eq:Hcd})),
\begin{align}
\dot{H}_\mathrm{CD}(t)&= i \hbar \sum_l | \ddot{l} \rangle \langle l | + | \dot{l} \rangle \langle \dot{l} | - \partial_t (\langle l | \dot{l} \rangle | l \rangle \langle l |) \nonumber \\
& = \,i \hbar \sum_l \dfrac{1}{2} (\vert \ddot{l} \rangle \langle l | - | l \rangle \langle \ddot{l} |) - \partial_t (\langle l | \dot{l} \rangle) | l \rangle \langle l | \nonumber \\
&- \langle l | \dot{l} \rangle (| \dot{l} \rangle \langle l | + | l \rangle \langle \dot{l} |).
\label{eq_Hcddot}
\end{align}
Finally, we can use Eq.(\ref{eq_Hcddot}) to evaluate  $\mathcal{P}_{CD}$ as follows\cite{mahunta2025shortcuts}, 

\begin{align*}
    \mathcal{P}_{\rm CD} &=  i\sum_l \lambda_l \left[\dfrac{1}{2} \left(\langle l \ddot{l} \rangle - \langle \ddot{l} | l \rangle \right) - \partial_t \Phi_{ll}-\Phi_{ll}  \partial_t(\langle l | l \rangle) \right] \non \\
    & = \;\dfrac{i }{2} \sum_l \lambda_l \left[ \langle l | \ddot{l} \rangle - \langle \ddot{l} | l \rangle - 2 \partial_t \Phi_{ll} \right] \nonumber \\
    &= \dfrac{i}{2} \sum_l \lambda_l \left[ \langle l | \ddot{l} \rangle - \langle \ddot{l} | l \rangle - \langle \dot{l} | \dot{l} \rangle - \langle l | \ddot{l} \rangle + \langle \ddot{l} | l \rangle + \langle \dot{l} | \dot{l} \rangle \right] \non\\
    &=0
\end{align*}

\section{Real-space structure of the non-unitary CD terms}
\label{Appreal}
In this appendix we analyze the real-space structure of the dissipative control terms and examine how LRIs affect their spatial profile. Starting from the momentum-space master equation [Eq.~\eqref{eq:master}], we focus on a representative contribution of the form $\sum_k \gamma_{12}^{(k)} (A_{12}^{(k)})^\dagger A_{12}^{(k)} \rho_k$. This generates terms such as
\begin{equation}
T_1^{(12)} = \sum_k \mathcal{K}^{(12)}_{1k} \, c_k^\dagger c_{-k} c_k c_k^\dagger c_{-k}^\dagger c_k \, \rho_k,
\end{equation}
where the kernel $\mathcal{K}^{(12)}_{1k}$ is given by  (see Eq.(\ref{Eq:population}))
\ba
\mathcal{K}^{(12)}_{1k} &=& \frac{\gamma_{12}^{(k)} |\phi^{(k)}_t|^2}{|\phi_t^{(k)}|^2 + 1} \non\\
&=&  \gamma_{12}^{(k)} \frac{f^2_{\alpha}(k)}{ (\mu +\cos(k) + E_k )^2 + f^2_\alpha(k)}.
\label{eq:K12}
\ea
Upon Fourier transforming to real space, the above term takes the form
\begin{equation} \label{eq:t12}
T_1^{(12)} = \frac{1}{L^3} \sum_k \mathcal{K}^{(12)}_{1k}
\sum_{n,m,p,q,r,s}
c_n^\dagger c_m c_p c_q^\dagger c_r^\dagger c_s \,
e^{i \mathcal{D} k} \, \rho,
\end{equation}
where $\mathcal{D} = ( n + m+ q ) - ( p + r+s) $,   represents an effective spatial separation between fermionic operators. Here we assume $\{n,m,p,q,r,s\}$ such that  $n > p > r > m > q > s$.  For a constant $\mathcal{D}$, and  in the thermodynamic limit, Eq. (\ref{eq:t12}) yields
\begin{equation}
T_1^{(12)} = \mathcal{J}_{12}(\mathcal{D}) 
\sum_{n,m,p,q,r,s}
c_n^\dagger c_m c_p c_q^\dagger c_r^\dagger c_s \, \rho,
\end{equation}
with the real-space control strength given by 
\begin{equation}
\mathcal{J}_{12}(\mathcal{D}) 
\approx \frac{2}{\pi} \int_0^\pi \mathcal{K}^{(12)}_{1k} \cos(\mathcal{D} k)\, dk.
\label{eq:J12}
\end{equation} 
% This expression makes it clear that the spatial structure of the dissipative control is governed by the momentum dependence of $\mathcal{K}_{1k}^{(12)}$, which in turn inherits the spectral properties of the system through $E_k$.
\begin{figure*}
    \centering
    \includegraphics[width=01\linewidth]{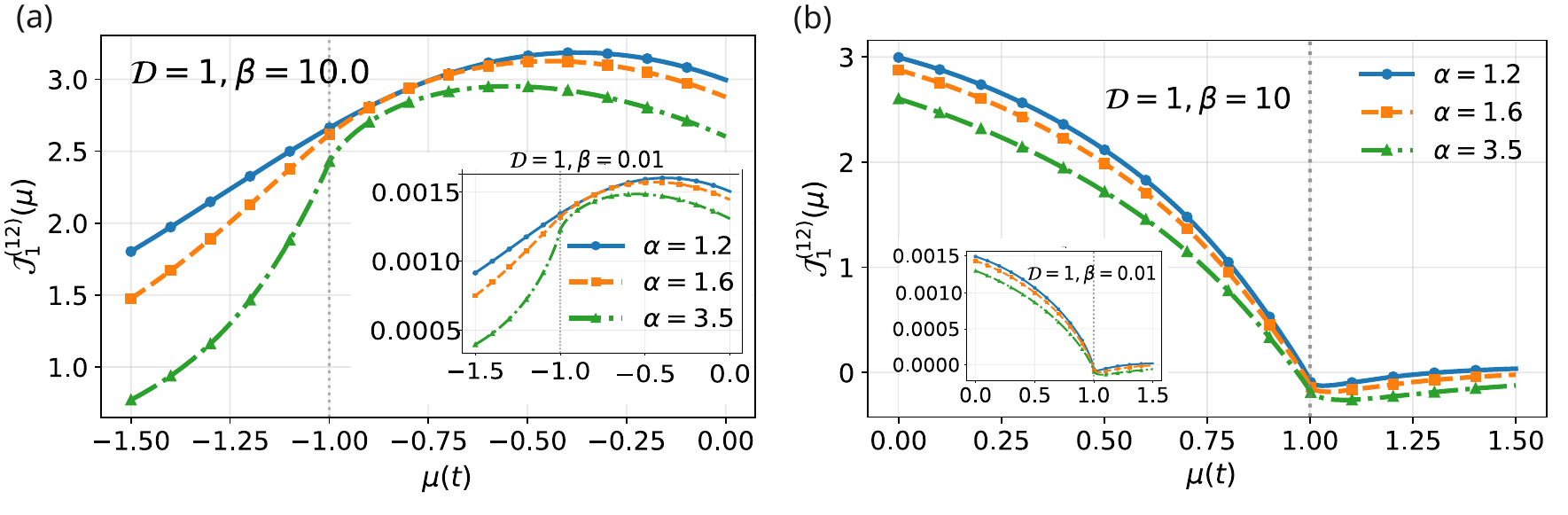}
    \caption{(a) Shows variation of the non-unitary control strength $\mathcal{J}_{12}$ as the chemical potential $\mu(t)$ is ramped across the critical point $\mu=-1$, for different values of the long-range exponent $\alpha$, at fixed $D=1$, in the low (main plot) and high (inset) temperature regimes, with $\beta=10$ and $\beta=0.01$, respectively.
(b) Evolution of $\mathcal{J}_{12}$ as $\mu(t)$ is swept across the critical point $\mu=+1$, at fixed $D=1$, in the low (main plot) and high (inset) temperature regimes, with $\beta=10$ and $\beta=0.01$, respectively.}
    \label{fig:J12_sweep}
\end{figure*}
The terms on the r.h.s. of Eq. \eqref{eq:J12} can be expected to vanish in the limit of large $\mathcal{D}$, as also verified numerically in Fig. \ref{fig:J12_with_D}, thus implying one can restrict $\mathcal{D}_1^{(12)}$ to small values only. 
Fig.~(\ref{fig:J12_sweep}) illustrates the behavior of the control strength $\mathcal{J}_{12}$ as a function of the distance from criticality for different values of $\alpha$, for a fixed $\mathcal{D}$ ($=1$). As shown in Fig.~(\ref{fig:J12_sweep}), $\mathcal{J}_{12}$ increases with increasing range of interactions (i.e., decreasing $\alpha$). Notably, the effect of $\alpha$ is suppressed close to criticality, which can be attributed to converging values of $\gamma_{12}$ for different values of $\alpha$, at the critical points (see Figs. \ref{fig:gamma_low_highTemp}a, \ref{fig:gamma_low_highTemp}e, and Eq. \eqref{eq:K12}). The insets of Fig.~ (\ref{fig:J12_sweep}) show that, in the high-temperature regime, $\mathcal{J}_{12}$ is significantly reduced, which is consistent with the fact that thermal mixing suppresses population gradients, thereby reducing the need for strong dissipative control.

%%%%%%%%

%\twocolumngrid 

%\bibliography{Refs.bib}

%% bbl starts here

\end{document}